\documentclass[fleqn]{article}
\usepackage{amssymb,amsmath}
\usepackage{amssymb,amsmath}
\usepackage[mathscr]{euscript}

\newcommand {\ignore} [1] {}

\newcommand{\qed}{\mbox{$\mathfrak{q.e.d}$}}
\def\newtheorems{\newtheorem{theorem}{Theorem}[section]
\newtheorem{cor}[theorem]{Corollary}

\newtheorem{lemma}[theorem]{Lemma}
\newtheorem{claim}[theorem]{Claim}

\newtheorem{definition}[theorem]{Definition}

}

\newtheorems

\newcommand{\CONTAINS}{\mbox{\tt Contains}}
\newcommand{\ADD}{\mbox{\tt Add}}
\newcommand{\REMOVE}{\mbox{\tt Remove}}

\newcommand{\New}{\mbox{$\tt NEW$}}

\newcommand{\Data}{\mbox{\it Key }}
\newcommand{\Event}{\mbox{$\mathit Event$}}

\newcommand{\Add}{\mbox{$\mathit Add$}}

\newcommand{\Rem}{\mbox{$\mathit Rem$}}

\newcommand{\Cnt}{\mbox{$\mathit Cnt$}}

\newcommand{\kval}{\mbox{$\mathit kval$}}

\newcommand{\Val}{\kval}

\newcommand{\AD}{\mbox{$\tt Ad$}}
\newcommand{\RM}{\mbox{$\tt Rm$}}

\newcommand{\Remove}{\mbox{\it REMOVE}}

\newcommand{\entry}{\mbox{$\mathit entry$}}

\newcommand{\Op}{\mbox{$\mathrm {Op}$}}
\newcommand{\End}{\mbox{$\mathit end$}}
\newcommand{\Begin}{\mbox{$\mathit begin$}}

\newcommand{\PMarked}{\mbox{\it Marked}}
\newcommand{\PActive}{\mbox{\it Active}}
\newcommand{\PNext}{\mbox{\it Next}}

\newcommand{\locate}{\mbox{\it Locate}}

\newcommand{\removal}{\mbox{\it removal}}

\newcommand{\checkit}{\mbox{\it CONTAINS}}
\newcommand{\Contain}{\mbox{\it CONTAINS}}

\newcommand{\InvA}{Inv\_A}
\newcommand{\InvB}{Inv\_B}
\newcommand{\InvC}{Inv\_C}
\newcommand{\InvD}{Inv\_D}

\newcommand{\PLockedto}{\mbox{$\mathit Locked\_to$}}

\newcommand{\PRemoved}{\mbox{$\mathit PhyRemoved$}}

\newcommand{\precc}{<}

\newcommand{\Address}{\mbox{\rm Address}}
\newcommand{\A}{\mbox{\rm Address}}
\newcommand{\adr}{\mbox{\rm adr}}
\newcommand{\Number}{\mbox{\rm Number}}

\newcommand{\activation}{\mbox{$\mathit activation$}}

\renewcommand{\L}{\mbox{$L_{LS}$}}

\newcommand{\pred}{\mbox{$\mathit pred$}}

\newcommand{\lock}{\mbox{$\mathit lock()$}}
\newcommand{\curr}{\mbox{$\mathit curr$}}
\newcommand{\next}{\mbox{$\mathit next$}}

\newcommand{\Nat}{\mbox{$\mathbb{N}$}}

\newcommand{\IF}{\mbox{\tt{if}}}
\newcommand{\THEN}{\mbox{\tt{then}}}
\newcommand{\WHILE}{\mbox{\tt{while}}}
\newcommand{\ELSE}{\mbox{\tt{else}}}

\newcommand{\goto}{\mbox{\tt{goto}}}

\newcommand{\Repeat}{\mbox{\tt{repeat}}}
\newcommand{\unlock}{\mbox{\tt unlock}}
\newcommand{\return}{\mbox{\tt return}}
\newcommand{\until}{\mbox{\tt until}}

\newcommand{\marked}{\mbox{$\mathit marked$}}

\newcommand{\Failed}{\mbox{$\mathit Failed$}}
\newcommand{\TRUE}{\mbox{$\mathit true$}}
\newcommand{\FALSE}{\mbox{$\mathit false$}}

\newcommand{\head}{\mbox{$\mathit Head$}}
\newcommand{\HH}{\head}

\newcommand{\Tail}{\mbox{\it Tail}}
\newcommand{\T}{\Tail}

\newcommand{\s}{\mbox{\it status}}

\newcommand{\p}{\mbox{\it pre}}
\newcommand{\cu}{c}

\newcommand{\LSstate }{\mbox{$L_{LSstate}$}}
\newcommand{\SimplerLSstate}{\mbox{$L_{SimplerLSstate}$}}

\begin{document}

\title{On the Lazy Set object}
\author{Uri Abraham
\thanks{Departments of Mathematics and Computer Science, Ben-Gurion
University.}}

\maketitle

\begin{abstract}
The aim of this article is to employ the Lazy Set algorithm as an example for a mathematical framework for proving the linearizability of a distributed system. The proof in this approach is divided into two stages of lower and higher abstraction level. At the higher level a list of ``axioms'' is formulated and a proof is given that any model theoretic structure that satisfies these axioms is linearizable. At this
level the algorithm is not mentioned. At the lower level, a Simpler Lazy Set algorithm is described, and it is shown that any execution of this simpler algorithm generates a model of these axioms (and is therefore linearizable). Finally the linearization of the Lazy Set algorithm is obtained by proving that any of its executions has a {\em reduct} that is an execution of the Simpler algorithm. So the reduct executions are linearizable and this entails immediately linearizability of the Lazy Set algorithm itself. 
\end{abstract}

\section{Introduction}

The motivation for this paper came while reading Dongol and Derrick article \cite{DD}  whose aim is to compare the major methods for verifying linearizability, and to describe their advantages and limitations.
That article  concentrates on three case studies of linearizability proofs in the aim
of identifying and exemplifying  basic approaches in the literature. A prominent case among the three is the Lazy Set algorithm
\cite{Helleretal},  which is interesting (also) because it ``represents a class of algorithms that can only be verified by 
allowing an
operation to set the linearization point of another [operation], and its proof is therefore more involved''. It is from \cite{DD} that we learned about the 
Lazy Set algorithm  and its central place  in the literature of linearizability proofs.
The Lazy Set algorithm implements a set object that supports three operations: Add, Remove, and Contains. The algorithm
is short, but the simplicity of its Contains operation is
rather misleading. As \cite{DD} states 
``Despite the simplicity
of this operation, its verification introduces significant complexity in the proof 
methods, requiring the use of more advanced verification techniques''.  For us this set algorithm is interesting because it
presents an opportunity to describe an approach to linearizability  that is different from the proof methods presented
in \cite{DD} in that it does not seek to find linearization points, but operates mainly at a higher abstract level.
It is the aim of our paper to exemplify this approach in some details.

The paper is not short, and for the reader who wants to see with no further delay in what sense our approach avoids the burden of
defining linearization points of the Contains method, we suggest concentrating on Section \ref{Sec2} which is the heart of the paper.
The proof of linearizability is divided into three parts which can be read independently. 
(1) Section \ref{Sec2} contains three axioms and a proof that any
structure (in the model theoretic sense) for these axioms is linearizable. (2) Section \ref{S4} contains a simpler version of the Lazy Set algorithm and a proof of its linearizability which shows that every execution of the Simpler algorithm satisfies the three axioms
of Section \ref{S4}. This simpler algorithm allows us to present the main ideas of the full Lazy Set algorithm in a more direct and compact setting. Moreover, the linearizability of the full algorithm is reduced to the Simpler Lazy Set algorithm. (3) This is done in 
Section \ref{SLSA} which contains (a version of) the full Lazy Set algorithm and a proof that any execution of the full algorithm contains
a reduct that can be seen as an execution of the Simpler one.

\section{The linear case: two specification modes}
\label{STS}

A {\em Set} data-structure $D$ is an object that suports three operations: Add, Remove, and Contains. 
The data added, removed and searched for is identified by {\em keys} which are taken to be natural numbers.
In applications, each key (id number) is associated with some information, but as \cite{Set}, \cite{Hell1},
and \cite{Helleretal} do, we ignore this data field and take the key to represent the data item. Thus we take $Data =\Data={\Nat}$.

The temporal precedence relation $<$ on the Add, Remove, and Contains operation executions is usually a partial ordering, and the
The Herlihy and Wing paradigm \cite{HW} for concurrent data-structure correctness is to first specify the data-structure when its operation executions
are linearly ordered in time (we use $<_0$ for this linear ordering), and then a distributed and concurrent execution is said to be
 linearizable if its precedence relation $<$ (typically a non-linear ordering) can be extended to  a linear ordering $<_0$ 
 which is correct
 according to the linearly ordered specification of the data-structure.
Our aim in this section is to define mathematically the collection of linearly ordered Set data-structures. We shall present here two
slightly different, but equivalent, modes of specifying this collection. The first  mode would seem more natural to most reader, we
call it the state-based mode, and the second mode turns out to be better suited for the proof approach presented here, and we call it
the functional mode.
In this section we henceforth assume that the operation executions are linearly ordered: for every two operation executions $e_1$ and
$e_2$ either $e_1<_0 e_2$ or $e_2<_0 e_1$. 

A process can invoke procedure $\ADD(v)$ with some parameter $v\in \Data$ when
 it seeks to add $v$ to $D$; it invokes $\REMOVE(v)$ in order to remove $v$ from $D$; and it invokes $\CONTAINS(v)$ to answer the question if $v$ is in $D$ or not. The parameter $v$ of each execution, $a$, of these operations is denoted $\kval(a)$, and when an operation terminates it returns a value   $\chi(a)\in \{ 0,1,f\}$, called the status of $a$,  whose meaning is explained now. Processes never fail stop and any execution
of an operation on $D$ terminates.  Yet, any Add or Remove  operation execution may fall short of its purpose, not in the sense that its processor breaks or stops, but
rather in that it fails to execute its mission, and in this case the operation returns the value $f$. This is a normal possibility,
caused by timing issues connected with interleaved operations by the other processes. A failed operation does not change the state of $D$, and the executing process can try again this (or any other)
operation. However, a $\CONTAINS(v)$ operation, and this is a key property of the Lazy Set algorithm, never fails and always returns
$1$ or $0$ (to say that $v$ is or is not in $D$).

If $e$ is an event then we use the predicates $\Add(e)$, $\Rem(e)$, and $\Cnt(e)$ to qualify $e$.
 If $\chi(e)\neq f$, then there are two possibilities.
 $\chi(e)=1$ indicates 
that when $e$ begins its execution $v=\kval(e)$ is in $D$, and $\chi(e)=0$ indicates that $v$ is not in $D$ at that moment. 
So, if $e$ is an \Add\
operation and $\chi(e)=1$, then $e$ does not change the state of $D$
 (for $D$ cannot contains two copies of $v$, it is a set not a multiset), but if $\chi(e)=0$ then the state of $D$ has changed
from one which does not contain $v$ to one that does contain that value. Correspondingly, if $e$ is a \Rem\ operation, then
$\chi(e)=0$ indicates that the set $D$ did not contain $v$ before the operation and hence that its state did not change, and
$\chi(e)=1$ indicates that the \Rem\ operation has changed $D$ from a state that contains $v$ to one that does not contain this
value.  If $\Cnt(e)$, then $\chi(e)=1$ indicates that $v=\kval(e)$ is in $D$, and $\chi(e)=0$ indicates that it is not.

The setting for the two specification modes is the same. We assume an infinite sequence of events $(e_i \mid i\in \omega)$ that
are linearly ordered by the temporal precedence relation $<_0$: $e_i <_0 e_j$ iff $i<j$; we  have three predicates that partition the set of events: \Add, \Rem, and \Cnt; and, in addition, two functions $\kval$ and $\chi$ are defined on the events. The intuitive meaning
of these predicates and functions is described as follows. For every
event variable $a$ (its value is some event $e_i$):
\begin{enumerate}
\item  $\Add(a)$, $\Rem(a)$, and $\Cnt(a)$ are formulas meant to describe the event type of $a$.
\item  $\kval(a)\in \Data$ points to the key value (in $\Nat$) which event $a$ is about to add, remove, or check.
\item  $\chi(a)\in \{0,1,f\}$ characterizes the status of $a$.  If $\Cnt(a)$ then $\chi(a)\in \{0,1\}$, and if $\Add(a)\vee \Rem(a)$
then $\chi(a)\in \{0,1,f \}$. $\chi(a)=f$ means failure of the event $a$ to terminate in a meaningful way. In this case $a$ has
no impact on the system. $\chi(a)=1$ means that $\kval(a)$ is already in the set being implemented, and $\chi(a)=0$ means
that it is not there.
 
\end{enumerate}

The two modes of specification give different but equivalent  answers to  the same question:
\begin{equation}
\label{EqSt}
\begin{minipage}[t]{125mm}
\begin{quote} {\em Given a sequence of events $(e_i \mid i\in \omega)$ with predicate  \Add, \Rem, \Cnt, and functions 
$\kval$, $\chi$ as described above, is this sequence representing a correct linear execution of a set data-structure?}\end{quote}
\end{minipage}
\end{equation}

\subsubsection*{The state mode specification}

If $D\subseteq \Data$ is a finite subset of $\Data$ then we say that $D$ is a {\em set state}.
 Given a sequence of events $(e_i \mid i\in \omega)$ with predicates and functions
as above, the state mode specification declares it to be a correct linear execution if there exists a sequence of 
set states $(D_i\mid i\in \omega)$ such that the following holds.
\begin{enumerate}
\item $D_i=\emptyset$. That is, initially the set is empty.
\item For every $i\in \omega$, the triple $(D_i, e_i,D_{i+1})$, together with $\kval(e_i)\in \Data$ and $\chi(e_i)\in\{0,1,f\}$
``correctly describes'' the effect of $e_i$ on the set state $D_i$ which results in state $D_{i+1}$.
\end{enumerate}
The detailed description of what ``correctly describes'' means is in the following table in which $a$ is an event variable
 (some $e_i$), $D$ is the state of tfhe set before $a$, and $D'$ is its state after $a$, so that the table describes correct
triples $(D,a,D')$ assuming that $x=\kval(a)\in \Data$. In case $\chi(a)=f$, then $a$ has to be an \Add\ or \Rem\ event, and
$D=D'$. Formula $\Add^p(a)$, for $p=0,1,f$ is a shorthand for the
conjunction $\Add(a)\wedge \chi(a)=p$, and $\Rem^p(a)$ is defined similarly. 
$\Cnt^p(a)$, however, is defined only for $p=0,1$.

\begin{table}[h!]
\def\arraystretch{1.3}
\centering
 \begin{tabular}{||c | c | c ||} 
 \hline
  event $a$ is s.t:  & condition on $x=\kval(a)$ and $D$ & condition on $D'$  \\ [0.5ex] 
 \hline \hline 

 $\Add^0(a)$ & $x\not\in D$ & $D'=D\cup\{x\}$  \\  [0.5ex] 
 $\Add^1(a)$ & $x\in D$ & $D'=D$  \\
 $\Rem^0(a)$ & $x\not\in D$ & $D'=D$  \\
 $\Rem^1(a)$ & $x\in D$ & $D'=D\setminus \{x\}$  \\
$\Cnt^0(a)$ & $x\not\in D$ & $D'=D$  \\
 $\Cnt^1(a)$ & $x\in D$ & $D'=D$  \\ 
$\chi(a)=f$   &  \text{no condition} & $D'=D$\\ [1ex] 
 \hline
 \end{tabular}
\caption{Description of the relation $(D,a,D')$.}
\label{T1}
\end{table}
We see in this table that there are only two cases when $a$ changes the state of $D$. When (1) $a$ is an \Add\ event and $\chi(a)=0$;
this is the case when $x=\kval(a)$ is not in the set $D$ just before $a$, and is added to $D$ to form $D'$ by $a$'s operation, and (2)
when $a$ is a \Rem\ event and $\chi(a)=1$; this is the case when $x=\kval(a)$ is in $D$ when $a$ begins its execution and is removed
from $D$ resulting in $D'=D\setminus\{x\}$. 

\subsubsection{The functional mode specification}
\label{SSF}
Although this state-based specification of the linear Set object is very natural and is surely the first that comes to mind, it turns out
that when we want to prove that a specific algorithm implements a set object,
 then (at least the way our
proof goes) the functional specification mode is handier. In this mode, the linear specification is based on a function $\gamma$ from a subset of the events into
the events. The specification is expressed with three sentences FS0, FS1 and FS2 that describe the properties
of $\gamma$. We need the following shorthands. For $p=0,1$, $\Op^p(a)$ is a shorthand for the formula
\begin{equation}
\label{EOp}
 (\Add(a)\vee\Rem(a)\vee\Cnt(a))\wedge \chi(a)=p.
\end{equation}
In addition we use the predicate $<_0$ which is assumed to be a linear ordering of the events. In the framework assumed above, $a_i<_0 a_j$
means simply that $i<j$.

\begin{figure}
\fbox{
\begin{minipage}[t]{125mm}

\begin{description}

\item[FS0:] $<_0$ is a linear ordering of the events. Predicates \Add, \Rem, and \Cnt\ are disjoint. The function $\gamma$
is defined over the set of $\Op^1$ events and returns $\Add^0$ events.
\item[FS1:] For every event $a$, if $\Op^1(a)$ then
\begin{equation}
\label{E1}
\begin{aligned} 
\gamma(a)<_0 a \wedge \Add^0(\gamma(a))\wedge \kval(a)=\kval(\gamma(a)) \wedge      \\
\neg \exists r(\Rem^1(r) \wedge \gamma(r)=\gamma(a) \wedge \gamma(a)<_0  r <_0 a).
 \end{aligned}
\end{equation}
\item[FS2:] For every events $a$ and $b$ such that $a<_0 b$, if $\Add^0(a)$ and $\Op^0(b)$ then:
\begin{equation}
\label{E2}
 \kval(a) =\kval(b) \to \exists r (a<_0 r <_0 b \wedge \Rem^1(r)\wedge  a=\gamma(r). 
\end{equation}
\end{description}
\end{minipage}
}
\caption{The Functional-Set serial specification is given by listing the properties of the $\gamma$ function. 
}
\label{F1}
\end{figure}

The Functional-Set specification is in Figure \ref{F1}. Assuming these properties we have the following lemma.

\begin{lemma}
For every event $x$ such that $\Op^1(x)$ there is no $\Op^0$ event $y$ such that
$\gamma(x)<_0 y<_0 x$ and $\kval(x)=\kval(y)$.
\end{lemma}
Proof. Assume on the contrary the existence of  $y$ such that $\Op^0(y)$ and $\gamma(x)<_0 y<_0 x$ and $\kval(x)=\kval(y)$.
 Apply FS2 to $\gamma(x)<_0 y$,
and get $r$ such that $\gamma(x) <_0 r <_0 y$,  $\Rem^1(r)$ and $\gamma(x)=\gamma(r)$, in contradiction to FS1. \qed

\begin{lemma}
\label{Lem4}
In Formula {\bf FS1} of Figure \ref{F1}, the second line of (\ref{E1})  is equivalent to the seemingly stronger
\begin{equation}
\label{Eq4}
\neg \exists b(\Rem(b) \wedge \chi(b)\neq f \wedge \gamma(a)<_0 b<_0 a \wedge \kval(b)=\kval(a)).  
\end{equation}
\end{lemma}
Proof. Assume that the function $\gamma$ satisfies the properties of Figure \ref{F1}. Suppose that there is an event $a$ such that
$\Op^1(a)$ and yet for some event $b$, 
\[ \Rem(b) \wedge \chi(b)\neq f \wedge \gamma(a)<_0 b<_0 a \wedge \kval(b)=\kval(a).\]
By the previous lemma it is not the case that $\Op^0(b)$, and hence $\Rem^1(b)$. By FS1 it is not the case that $\gamma(b)=\gamma(a)$,
and hence either $\gamma(a)<\gamma(b)$ or else $\gamma(b)<\gamma(a)$. But both alternatives contradict the previous lemma.
The first alternative contradicts the lemma when applied to $x=a$, $y=\gamma(b)$, and the second when applied to $x=b$, $y=\gamma(a)$. \qed

 The equivalence of the two linear specifications is expressed in the following.
\begin{theorem}
The state-based and functional-based specifications are equivalent in the following precise sense.
If $(e_i\mid i\in \omega)$ is any sequence with predicates and functions as in the framework presented above in (\ref{EqSt}), then the following
hold.

\begin{enumerate}
\item If there is a sequence of set states $D_i\subseteq \Data$ for $i\in\Nat$ that shows that the given sequence of events satisfies the state-based
specification, then there is a function $\gamma$ on the $\Op^1$ events in the sequence which satisfies properties FS0, FS1 and FS2
of Figure \ref{F1}.
\item If There is a function $\gamma$ on the $\Op^1$ events in the sequence $(e_i \mid i\in \omega)$ which satisfies properties FS0, FS1 and FS2, then there exists
a sequence $(D_i\mid i\in\omega)$ of set states so that $D_0=\emptyset$ and each triple $D_i, e_i,D_{i+1}$ complies with the
requirements of Table \ref{T1}.
\end{enumerate}
\end{theorem}
Proof. For the first item of the theorem, assume an evidence sequence $(D_i\mid i\in\Nat)$ for the state based specification. Assume that event $e_i$ is
such that $\Op^1(e_i)$, and say  $x=\kval(e_i)$. We have to define $\gamma(e_i)$. Consider the triple $(D_i,e_i,D_{i+1})$ and lines 2,4, and 6 in Table \ref{T1}. 
Since $\chi(e_i)=1$, $x\in D_i$ follows. Since 
$D_0=\emptyset$, there is an index $k<i$ such that $x\not\in D_k$. Let $k<i$ be the maximal such index, and then for every
$j$ such that $k+1\leq j \leq i$, $x\in D_j$. We define $\gamma(e_i)=e_k$. Since $x\not\in D_k$ but $x\in D_{k+1}$, $\Add^0(e_k)$
and $\kval(e_k)=x$. Clearly, by the maximality of $k$, there is no index $j$ with $k<j<i$ and such that $\Rem(e_j)\wedge \kval(e_j)=x$.
Thus FS1 holds.

 To prove FS2, assume that $\Op^0(e_i)$, and $a=e_j<e_i$ is such that $\Add(a)\wedge \kval(a)=x=\kval(e_i)$. 
So $x\in D_{j+1}$. This indicates that $j+1<i$ and there has to be some index $k+1\leq m <i$ such that
 $\Rem (e_m)$ and $\kval(e_m)=x$ as required. 

For the second item of the theorem, given a series of events $(e_i\mid i\in\omega)$,  assume a function $\chi$ with values in $\{0,1,f\}$, and a function $\gamma$ that is defined on the $\Op^1$ events in the sequence and is such that properties of Figure \ref{F1}  hold. 
We define a sequence
of sets $D_i\subseteq \Data$ by induction on $i$, and then prove that this sequence is an evidence that the sequence of events
satisfies the linear state specification summarized in Table \ref{T1}. (The definition of $D_i$ does not depend on $\gamma$, but
its properties do.) For $i=0$ define $D_0=\emptyset$. Assuming that $D_i$ is defined consider event $e_i$ and
say $ k =\kval(e_i)$. If $\chi(e_i)=f$ then $D_{i+1}=D_i$. Assume now that $\chi(e_i)\neq f$.
\begin{enumerate}
\item If $\Add(e_i)$, then define $D_{i+1} = D_i\cup \{ k\}$. (So if already $k\in D_i$ then $D_{i+1}=D_i$).
\item If $\Rem(e_i)$, then define $D_{i+1}=D_i\setminus \{k\}$. (So if $k\not\in D_i$ then $D_{i+1}=D_i$).
\item If $\Cnt(e_i)$, then define $D_{i+1}=D_i$.
\end{enumerate}
It is evident that we had no choice but to define the $D_i$ sequence in this way, but this does not mean that the sequence is
a good evidence for the state specification. What we have to prove is that each triple $D_i,e_i,D_{i+1}$ satisfies the requirements
of Table \ref{T1}. Essentially this means that if $k=\kval(e_i)$ and $\chi(e_i)\neq f$, then 
\begin{equation}
\label{Eq5}
 k\in D_i \text{ iff }\chi(e_i)=1.\end{equation}
This claim is proved by induction on $i$. Assuming (\ref{Eq5}) holds for all indexes below $i$ we prove the claim for $i$. 
So assuming that $k=\kval(e_i)$ and $\chi(e_i)\neq f$ there are two cases:
\begin{enumerate}
\item Case $\chi(e_i)=1$. Then $\Op^1(e_i)$ and FS1 applies. So there is $e_j<e_i$ (namely $e_j=\gamma(e_i)$) such that
 $\Add^0(e_j)$, $\kval(e_j)=k$ and there is no $\Rem^1$ event $r$ between $e_j$ and $e_i$ such that 
\[ \Rem(r)\wedge \chi(r)\neq f \wedge \kval(r)=\kval(e_i)\] (by Lemma \ref{Lem4}). This implies that 
$k$ ``remains''
in each $D_n$ for $j < n \leq i$, and hence that $k\in D_i$.

\item Case $\chi(e_i)=0$. We must prove that $k\not\in D_i$. Suppose on the contrary that $k\in D_i$.
Since $D_0=\emptyset$ there has to be some maximal $j$ such that $j\leq i$ and $k\not \in D_j$.  So $j<i$, $k\not\in D_j$ and
$k\in D_{j+1}$. Hence $\Add(e_j)$ and $\kval(e_j)=k$. By the induction assumption
$\chi(e_j)=0$. I.e. $\Add^0(e_j)$.
 
 By FS2 there has to be some
$r=e_n$ such that $e_j<e_n<e_i$ and $\Rem^1(e_n)\wedge \kval(e_n)=k$. This implies that $k\not\in D_{n+1}$.
This contradicts the maximality of $j$.
\end{enumerate}

\section{Axioms for the general non-linear case}
\label{Sec2}

Having specified the Set object in the linear case, the linearization approach of Herlihy  and Wing \cite{HW} immediately yields
the specification of the Set object in arbitrary distributed executions, namely by requiring that the temporal precedence ordering $<$
on the events 
(a partial
ordering) has an extension $<_0$ to a linear ordering that satisfies the linear specification of Figure \ref{F1}. 
 Although 
linearizability clarifies {\em what} has to be proven, it does not indicate {\em how} it is to be proven, and this remains difficult in many cases.
 The abstract properties 
listed in Figure \ref{NL1} form an intermediary stage between linearizability as a definition of correctness and the correctness proof itself. In our experience, this isolation of abstract properties (exemplified here with the Lazy Set axioms) helps in finding correctness
proofs of nontrivial distributed algorithms and in presenting these proofs in a clear way. The correctness proof is divided in two stages,
a proof that the abstract properties imply linearizability and a proof that any execution of the algorithm satisfies the abstract properties.
In the first stage we are not bothered by the details of the algorithm and in the second we have a limited (and thus simpler) aim, namely a proof of the abstract properties rather than a direct proof of linearization.

\subsection{Tarskian system executions}
   \label{S3.1}

The term ``structure'' has such a wide range of meanings that we must be clearer and say that in this article 
we use this term as it is defined in mathematical logic textbooks: a structure is an {\em interpretation}
of a predicate logic language. We may even write  ``Tarskian structures'' and ``Tarskian system-executions'' for this specific
usage.  Readers who are not familiar with this concept will find that the following self-contained
 discussion of the Lazy Set language and its interpretations suffices for reading the linearizability proof presented here.
 
\begin{definition}
\label{DLang}
The Lazy Set language $\L$ is a multi-sorted language. It has two sorts (types of elements) {\em Event} and {\em natural-number}. 
  Additional symbols in $\L$ are the following predicates and function symbols. 
	\end{definition}
\begin{enumerate} 
\item The predicates (names of relations)
are the unary predicates: {\em low-level}, {\em high-level},  \Add, \Rem, and \Cnt\ that are defined over sort \Event, and the binary predicate
$<$ that is defined on sort \Event\ (and on sort natural-numbers). Also, every process name $P$ is used as a unary predicate.
\item There are five function symbols in $\L$.  $\Begin, \End:\Event\to \Event$,
$\chi:\Event\to \{0,1,f\}$, $\gamma:\Event\to \Event$, and $\kval:\Event\to \Nat$.
\end{enumerate}
Viewed as elements of the language these are just symbols, but their names reflect 
 certain {\em intentions} and preconceptions. However, in the proof of linearizability (Theorem \ref{MT}) this 
tacit intuitive understanding cannot be used, and only formal assumptions expressed in this language (the ``axioms''  displayed in Figure \ref{NL1}) count.
 Sort $\Event$ refers to  the set of events of the universe of discourse. A high level event is a set of low level events; usually
some operation execution. The lower level events (also called actions) are, for example, read and write actions. If $\Event(e)$ (i.e. $e$ is an event) 
then $\Cnt(e)$ for example ``says'' that $e$ is a {\em contains} event, 
$\kval(e)$ is the key value of that event, a natural number, and $\chi(e)\in\{0,1,f\}$ is the status of this
event. We can state that $e$ is a high level event by the formula {\em high-level}$(e)$. (For those who jumped directly here, reading of subsection \ref{SSF} can help to understand our choices for the language $\L$.)

What may seem strange when checking the statements at Figure \ref{NL1} is   that \Rem\ and \Add\ events are considered to be actions rather
than high level events.
This will become clearer later, but now we can say that while it is true that any \REMOVE\ and \ADD\ operation execution is a higher  
level event, each such operation execution contains a distinctive and easily identified ``linearization point'' which is an action
that can be identified with  the operation itself. So the intention is that \Add\ refers to the linearization actions of \ADD\  
operation executions, and \Rem\ refers to the linearization actions of \REMOVE\ operation executions. The predicate \Cnt, however, refers
to high level events, not to low level actions.

A Tarskian structure that interprets $\L$ consists of a universe $|M|=\Event^M\cup \mbox{natural-numbers}^M$ which is
the disjoint union of the interpretations of its sorts. Then each predicate $Q$ is interpreted by $M$ as a relation $Q^M$
over the universe of $M$ in a way that respects the sorts. For example, $\Add^M\subseteq \Event^M$, and $<^M$ is a binary
relation over $\Event^M$. Any function symbol $F$ is interpreted as a function over the universe of $M$ that respects the sorts.
For example, $\chi^M:\Event^M\to \{0,1,f\}$.

Even though we have in mind a certain intention for the symbols of $\L$, the interpreting structures do not have to respect it, and
$<^M$ for example can be interpreted as an arbitrary binary relation on the events. The term Tarskian
system-execution\footnote{We follow and extend the notion 'system execution' that was
introduced by Lamport \cite{Lamport}, and adapted to the model theoretic framework, for example, in \cite{A1} and \cite{A2}.} however refers to  interpretations of this language  that respect an additional restrictions: 
that relation $<^M$ is a partial ordering with the property that every event has only a finite number of predecessors. 
In fact, a stronger property ({\em Lamport's finiteness property}) holds in system executions.
\begin{equation}
\label{LFT}
\begin{aligned}
&\text{For every event } x \text{ there exists a finite set } E \text{ such that}\\
&\text{for every event } y, 
y\not\in E \Rightarrow x<y.
\end{aligned}
\end{equation}
The following Lazy Set axioms are specific restrictions imposed by sentences in $\L$.

\subsection*{The Lazy Set Axioms}
Figure \ref{NL1} presents   properties (the Lazy Set ``axioms'') which are (English rendering of) sentences written in the logical 
language $\L$ described above. For example, the statement in A0 that ``The \Add,\Rem, and \Cnt\ predicates  on the events 
are pairwise disjoint''
can be rendered as \[\neg \exists e ((\Add(e)\wedge \Rem(e))\vee (\Add(e)\wedge \Cnt(e))\vee(\Rem(e)\wedge\Cnt(e)).\]
And, the additional information that the \Add\ predicate for example is over the event sort can be rendered as 
$\forall e(\Add(e)\to \Event(e))$ (the name of a sort can be used as a predicate).

The following informal discussion explains the meaning of these properties by referring to a 
system execution $M$
that interprets $\L$.
There are two kinds of events in $M$: low level events, also called actions, and high level events, also called operation executions.
A low level event can represent a write action on some memory register, a locking event etc. and a high level event is a finite set
of low level events that represents some operation execution. Predicates {\it low-level} and {\it high-level} are over the \Event\ sort, 
and  a ``membership'' predicate $\in$ can relate a low level event to the high level
event that contains it.  The low level events are assumed to be linearly
ordered by the temporal precedence relation $<$. This is stated in axiom A0; what is not stated in that axiom (it is a general
property of system executions, not specific to the axioms) is that every event has only a finite number of predecessors in $<$.

The functions \Begin\ and $\End$ are used to give to every high level event $X$ the first and last actions $\Begin(X)$ and
$\End(X)$ that $X$ contains. If is convenient to define $\Begin(e)=\End(e)=e$ for every low level event.

If $e$ is an event in $M$ and  $\Add(e)$ holds, then $e$ represents an event of adding the key value $\kval(e)$ to the set,
and $\chi(e)\in \{ 0,1,f\}$ is  the status of $e$ ($\chi(e)=0$ says that the key $\kval(e)$ was added, and $\chi(e)=1$ that it was not added because it was already there).

In system executions the (partial) ordering on the high-level events is assumed to be consistent with the linear ordering on the actions in the sense that
if $e_1$ and $e_2$ are high-level events then $e_1<e_2$ iff $\forall x\in e_1\ \forall y\in e_2 (x<y)$. Equivalently,
\[e_1<e_2 \text{ iff } \End(e_1)<\Begin(e_2).\]
We also define $a<e$ for an action $a$ and higher-level event $e$ iff $a<b$ for all $b\in e$. Equivalently, $a<e$ iff
$\End(a)<\Begin(e)$.  Similarly $e<a$ is defined
when $b<a$ for all $b\in e$. 

If action $a$ is not in $e$ then $a\not<e$ is equivalent to $\Begin(e) < a$.

\begin{figure}[ht!]
\fbox{
\begin{minipage}[ht]{130mm}

\begin{description}

\item[A0:] 
The \Add,\Rem, and \Cnt\ predicates  on the events are pairwise disjoint. The \Add\ and \Rem\ are
low level events (actions) and the \Cnt\ events are higher level events. For every event $X$, $\Begin(X)$ and
$\End(X)$ are actions. For \Add\ and \Rem\ events $E$, $\Begin(E)=\End(E)=E$. But for \Cnt\ events $E$, $\Begin(E)<\End(E)$. 
Restricted to the low level actions, $\precc$ is a linear ordering.

\item[A1:] 
For every $\Op^1$ event $A$:
\begin{equation}
\label{E4}
\begin{aligned} 
&\Add^0(\gamma(A))\wedge \kval(\gamma(A)) =\kval(A)\wedge \gamma(A) <\End(A) \wedge\\
& 
\neg \exists R ((\Rem^1(R)\wedge \gamma(R) = \gamma(A) \wedge \gamma(A)\precc R \precc A)
\end{aligned}
\end{equation}

\item[A2:]
For every events $A$ and $B$:
\begin{equation}
\label{E4a}
\begin{aligned}
&  \ \Op^0(B) \wedge  \Add^0(A)\wedge  A\precc B \wedge \kval(A)=\kval(B) \Rightarrow \\
&\exists R (\Rem^1(R)\wedge A=\gamma(R)\wedge R<\End(B)).
\end{aligned}
\end{equation}
\end{description}
\end{minipage}
}
\caption{The non-linear Lazy Set axioms. $\Op^0$ and $\Op^1$ are shorthand for formulas defined in equation (\ref{EOp}). }
\label{NL1}
\end{figure}
We now explain the three axioms and argue that they are natural properties implied by linearizability.

\noindent
\underline{Concerning A0.}
 It is noted already in \cite{DD}
(and in the papers that are surveyed  there) that the \Add\ and \Rem\ operation executions have natural and easily defined linearization points.
In fact, it is possible to point to specific atomic instructions in these protocols that serve in each $\Add^0$, $\Add^1$, $\Rem^0$, 
and $\Rem^1$ 
executions as linearization points.   
Assuming that this is indeed the case we can take these linearization points as representatives of the whole \Add\ and \Rem\ operation execution. Thus, the \Add\ and \Rem\ operation executions are represented in the Lazy Set axioms
as actions rather than as high level events. This is the meaning of the statement in A0 which says that
the \Add\ and \Rem\ events are linearly ordered actions.

\noindent
 \underline{Concerning A1.} Whereas the \Add\ and \Rem\ events are actions, the \Cnt\ operations are high level events, i.e. sets of actions delineated by their first and last actions
$\Begin(e)$ and $\End(e)$. The function $\chi$ is defined on the events and returns values in $\{0,1,f\}$. In the Lazy Set algorithm,
the $\chi$ value of \CONTAINS\ operation executions is never $f$, but there is no need to add this fact to the axioms. Note that nothing is
said about failed \Add\ and \Rem\
events in these axioms. (Of course, in proving that these axioms hold it is necessary to observe that failed events do not change the
state of the system.) The function $\gamma$ is defined on events $A$ such that $\chi(A)=1$
 and then $\gamma(A)$ is some $\Add^0$
event (that is $\chi(\gamma(A))=0)$. The function $\gamma$ already appears in the linear specification of the set object
in Figure \ref{F1} in property FS1 which says that
for every event $a$, if $\Op^1(a)$ then
\begin{equation}
\label{EE1}
\begin{aligned} 
\gamma(a)<_0 a \wedge \Add^0(\gamma(a))\wedge \kval(a)=\kval(\gamma(a)) \wedge      \\
\neg \exists r(\Rem^1(r) \wedge \gamma(r)=\gamma(a) \wedge \gamma(a)<_0 r <_0 a).
 \end{aligned}
\end{equation}
The only difference between FS1 and A1 is that $\gamma(A)<_0 A$ is replaced with the weaker $\gamma(A)<\End(A)$ (which makes
a difference only when $\Cnt(A)$ holds).
Clearly if even this weakening of FS1 does not hold and it happens that $\End(A)<\gamma(A)$, then in any linear augmentation $<_0$ of $<$ we would have that
$\End(A)<_0\gamma(A)$ and this would contradict linearizability.

\noindent
\underline{concerning A2.}    Since for every action $B$, $\End(B)=B$, A2 says about {\em actions} exactly
 what FS2 says. If however $\Cnt^0(B)$, then A2 replaces the existence of $\Rem^1$ event $r$ such that
$r<_0 b$ with the weaker requirement $r<\End(b)$.
It is easy to see that A2 cannot be replaced by a weaker condition.

Thus properties A0--A2 weem to be weaker than the linear Set specification FS0--FS2 of
Figure \ref{F1}, but they are in fact strong
enough to entail linearizability. That is, if $M$ is any Tarskian system-execution that satisfies the Lazy Set axioms
then there exists a linear extension $<_0$ of the precedence relation $<^M$ which satisfies the required linear Set
specification FS0--FS2. Of course, since
the abstract Lazy Set axioms do not mention any specific algorithm, a proof of linearizability in this context cannot rely on 
linearization points--it must remain on the abstract level. 
The following is our main theorem.

\begin{theorem}
\label{MT}
The Lazy Set axioms A0--A2 imply linearizability of any system execution for $L_{LS}$.
\end{theorem}
Proof. The proof relies on four lemmas and a theorem in which the
Lasy Set axioms
 are tacitly assumed.

\begin{lemma}
\label{L1}
If $Z$ is some $\Add^1$\ or $\Rem^1$ event then $\gamma(Z)< Z$.
\end{lemma}
Proof. By A1, $\gamma(Z) <\End(Z)$, but since  $Z$ is an action, $\End(Z)=Z$, so that
$\gamma(Z)<Z$ follows.
\qed

\begin{lemma}
\label{L2.3}
For every $\Op^1$ event $x$, there is no $\Op^0$ event $y$ such that \[\kval(y)=\kval(x) \wedge \gamma(x)<y<x.\]
\end{lemma}
Proof. Assume on the contrary the existence of such events $x$ and $y$. Apply A2 to $A=\gamma(x)$ and $B=y$, to get an event
$R$ such that $R<\End(y)$ and $\Rem^1(R) \wedge A=\gamma(R)$ (which implies $A<R$ by the previous lemma).
 Hence $R<x$ (since $y<x$). But now $\gamma(x)<R<x$
is in contradiction to A1. \qed 

We now define a relation $\Rightarrow$ on the events which turns out to be
 a key ingredient in the definition of the linear 
extension $<_0$ of $<$.

\begin{definition}
\label{DA}
The $\Rightarrow$ relation is defined in the following cases:
\end{definition}
\begin{enumerate}
\item For every $\Cnt^1$ event $C$, define 
\begin{equation}
\label{E5}
\gamma(C)\Rightarrow C,
\end{equation}
and if $R$ is such that  $\Rem^1(R)$ and $\gamma(R)=\gamma(C)$ then define
\begin{equation}
\label{E6}
C\Rightarrow R.
\end{equation}

\item For every events $C$ and $R$, if $\Cnt^0(C)$ and $\Rem^1(R)$, if $\kval(R)=\kval(C) \wedge  C\not< R  \wedge \gamma(R)\precc C$,
then we define 
\begin{equation}
\label{E7}
R \Rightarrow C.
\end{equation} 

\item For every events $C$ and $A$ with $\kval(C)=\kval(A)$, if $\Cnt^0(C)\wedge \Add^0(A)\wedge A \not < C$, then
\begin{equation}
\label{E8}
C\Rightarrow A.
\end{equation}
\end{enumerate}

\begin{lemma}
\label{L2.4}
For every events $X$ and $Y$, if $X\Rightarrow Y$ then $Y\not < X$ (equivalently $\Begin(X)<\End(Y)$).
\end{lemma}
Proof. We follow the cases in Definition \ref{DA} of the $\Rightarrow$ relation, and prove that $\neg(Y\precc X)$ holds whenever
$X\Rightarrow Y$.
\begin{enumerate}
\item In case $\gamma(C)\Rightarrow C$ where $\Cnt^1(C)$, the impossibility of $C\precc \gamma(C)$ is a consequence of A1 (which
implies that $\gamma(C)<\End(C)$).
And if $R$ is such that $\Rem^1(R)$ and $\gamma(R)=\gamma(C)$ (in which case we defined $C\Rightarrow R$) then the impossibility of $R\precc C$
is again a consequence (mainly) of A1. For if $R\precc C$ then $A=\gamma(C)=\gamma(R)$ is an $\Add^0$ event such that
$A=\gamma(C)\precc R \precc C$ and $\kval(R)=\kval(A)$ are in contradiction to A1. Why $A=\gamma(R)\precc R$? By Lemma \ref{L1}.

\item In case $R\Rightarrow C$ where $\Cnt^0(C)$ is as in item 2 of Definition \ref{DA}, then $\neg(C\precc R)$ is specifically required
in that definition.

\item Likewise, in case 3, in which $C\Rightarrow A$ is defined, $\neg(A\precc C)$ is explicitly required. 
\end{enumerate} 

\begin{lemma}\label{L2.5}
Suppose that $X\Rightarrow Y \Rightarrow Z$. 
\begin{enumerate}
\item If $\Add(X)$, then $X\precc Z$, $\Cnt^1(Y)$ and $\Rem^1(Z)$.
\item If $\Rem(X)$, then $X\precc Z$, .
\item If $\Cnt(X)$ then $\Begin(X)<\End(Z)$.

\end{enumerate}

\end{lemma}
Proof. Assume $X\Rightarrow Y \Rightarrow Z$, and check the three cases of the lemma.
\begin{enumerate}
\item  In case $X$ is an \Add\ event, we observe that there is only one case in the definition of $\Rightarrow$
in  which an \Add\ event is to the left of an
$\Rightarrow$ relation; this is item 1, in which $X=\gamma(C)$ for some $\Cnt^1$ event $C$. So the situation is
$X=\gamma(C)\Rightarrow C\Rightarrow Z$, and we want to prove that $\Rem^1(Z)$ and $X<Z$. Well,   
there is only one case in which a $\Cnt^1$ event $C$ is to the left of the
$\Rightarrow$ arrow, and it is again item 1: $C\Rightarrow R$ where $\Rem^1(R)$ and $\gamma(R)= \gamma(C)$.
Since $\gamma(R)=\gamma(C)=X$, Lemma \ref{L1} implies that $X\precc R$.
In the notation of item 1 of the lemma, this means that $X\precc Z$. 

\item Suppose now that $\Rem(X)$.  There is only one case in which a \Rem\ event $X$ is to the left 
of some $\Rightarrow$ relation. This is item 2 where $X\Rightarrow C$ are such that $\Rem^1(X)$,  $\Cnt^0(C)$, 
$\kval(X)=\kval(C)$, $C\not\precc X$, and $\gamma(X)\precc C$. So now $X\Rightarrow C\Rightarrow Z$. 
Since $\Cnt^0(C)$, relation $C\Rightarrow Z$ is only possible by item 3, and hence it has to be the case that $\Add^0(Z)$,
 $\kval(Z)=\kval(C)$, and $Z\not\precc C$.

Since $\gamma(X)\precc C$ but $Z\not\precc C$, and as $\gamma(X)$ and $Z$ are comparable, $\gamma(X)\precc Z$ follows (or else
$Z\leq \gamma(X)<C$ would entail $Z<C$). We want to
prove that $X\precc Z$. Assume that this is not the case and hence that $Z\precc X$ (use comparability of actions). So $\gamma(X)\precc Z\precc X$ where
$\Add^0(Z)$, $\Rem^1(X)$ and the three events $\gamma(X), Z,X$ have the same value.
But this contradicts Lemma \ref{L2.3}.

\item Now suppose that $\Cnt(C)$ and $C\Rightarrow Y\Rightarrow Z$. We have to prove that $\Begin(C)<\End(Z)$. There are two
cases here: $\Cnt^0(C)$ and $\Cnt^1(C)$.
\begin{enumerate}
\item Assume that $\Cnt^0(C)$. Then $C\Rightarrow Y$ implies that item 3 in the $\Rightarrow$ definition applies. That is, $\Add^0(Y)$,
$C$ and $Y$ have the same value, 
and $Y\not< C$. That is, $\Begin(C) <\End(Y)$. Since $Y\Rightarrow Z$, only item 1 applies, and hence $\Cnt^1(Z)$ and $Y=\gamma(Z)$. 
This implies
that $Y<\End(Z)$ (by A1), and hence $\Begin(C) < \End(Z)$ follows.

\item Assume that $\Cnt^1(C)$. Then $C\Rightarrow Y$ implies that item 1 in the $\Rightarrow$ definition applies.
So $\Rem^1(Y)$ and $\gamma(C)=\gamma(Y)$. By Lemma \ref{L2.4},  $\Begin(C)<\End(Y)$ which is the same as $\Begin(C)<\Begin(Y)$ 
(since $\End(Y)=\Begin(Y) =Y$
for any \Rem\ event $Y$). If in addition $Y\Rightarrow Z$ holds, then $Y=\Begin(Y)<\End(Z)$ by Lemma \ref{L2.4} again. Hence $\Begin(C) < \End(Z)$ in this case as well.

\end{enumerate}
\end{enumerate}
\qed

\begin{theorem}
Relation $\precc' = (\precc \cup \Rightarrow)$ contains no cycles.
\end{theorem}
Proof. By definition of union of relations, $a<' b$ iff $a<b$ or $a\Rightarrow b$.  
A cycle is a sequence $X_0\precc' X_1\precc' \cdots\precc' X_n$ where $X_0=X_n$. We say that $n$ is the {\em length } of this cycle.

 (1) There are no cycles of length 1. I.e. no cycles of the form $X \precc' X$. Indeed it can neither be the case that $X\precc X$
nor that $X\Rightarrow X$. 

(2)
 There are no cycles of length 2. I.e. there are no cycles of the form $X\precc'Y\precc' X$. Indeed, 
 $X\Rightarrow Y \precc X$ is impossible by Lemma \ref{L2.4} which deduces from $X\Rightarrow Y$ that it is not the case that $Y\precc X$. 
But $X\Rightarrow Y\Rightarrow X$ is also impossible, because the only possibilities in the $\Rightarrow$ relation are
$\Add^0\Rightarrow \Cnt^1$, $\Cnt^1\Rightarrow \Rem^1$, $\Rem^1\Rightarrow \Cnt^0$, and $\Cnt^0\Rightarrow \Add^0$.

(3) There is no cycle of length $3$. Suppose on the contrary that $S \precc' T \precc'  U\precc' S$ is such a cycle. 
There are no two occurrences of the $\precc$ relation on that cycle, or else it could be shortened by transitivity of $<$.
 Hence there are two
occurrences of the $\Rightarrow$ relation and we may assume that the cycle begins with these two occurrences:
$S\Rightarrow T \Rightarrow U \precc' S$. We have two possibilities.

\begin{enumerate}
\item   $S\Rightarrow T\Rightarrow U\precc S$. Then $\Begin(S)<\End(U)$ (by Lemma \ref{L2.5}) which contradicts $U\precc S$.

\item $S\Rightarrow T \Rightarrow U \Rightarrow S$. The cycle has to contain an \Add\ event (or else it would contain two successive
\Rem\ or two successive \Cnt\ events, but an inspection of the $\Rightarrow$ definition shows that this is never the case).
So we may assume that $S$ is an \Add\ event, and in this case $S\precc U$ by Lemma \ref{L2.5}. So $S<U\Rightarrow S$ which is
a cycle of length 2.

\end{enumerate}

(4) There is no cycle of length $\geq 4$. Suppose on the contrary that a cycle of length $\geq 4$ exists and consider one of
minimal length. 
\begin{enumerate}
\item[{Case 1}:] The cycle contains three successive appearances of the $\Rightarrow$ relation: 
$S\Rightarrow T\Rightarrow U\Rightarrow V$ (possibly $S=V$).  If $S$ is an \Add\ or \Rem\ event, then $S\precc U$ (Lemma \ref{L2.5}) which yields a shorter sequence.
If $S$ is a \Cnt\ event, then $T$ is an \Add\ or \Rem\ event, and then $T\precc V$ which yields again a shorter sequence. Thus we may dismiss Case 1.
\item[{Case 2}:] There are two successive appearances of the $\Rightarrow$ relation on the cycle but not three. Let $T\Rightarrow U \Rightarrow V$
be such a pair of successive relations. Let $S$ be that event on the cycle that appears before $T$, and let $W$  be that
event on the cycle that appears after $V$.
Then necessarily $S\precc T$ and $V\precc W$ (again because there are no three successive $\Rightarrow$ relations
on the cycle). So $S\precc T\Rightarrow U\Rightarrow V \precc W$. Note that $T\not = W$
(since that cycle is of length $\geq 4$).
Now $\Begin(T)<\End(V)$ by Lemma \ref{L2.5}, and hence $S\precc W$ follows. So $S\neq W$ and since $W$ is not $T$
(i.e. $W$ is not the successor of $T$ on the cycle) we get a shorter cycle by making a cycle in which $W$ is the successor of $T$.
This is in contradiction to the minimality of the cycle.  

\item[Case 3:] There are no two successive $\Rightarrow$ relations on the cycle. Hence the cycle alternates $<$ and $\Rightarrow$
relation. We may assume that it begins with $\precc$ and its first three relations are: $S\precc T \Rightarrow U \precc V$. Then $\Begin(T)<\End(U)$ and hence $S\precc V$ follows, and the cycle can be shortened. \qed
\end{enumerate}  

The proof of Theorem \ref{MT} can now be concluded. Since relation $\precc' =\precc\cup\Rightarrow$ has no cycles it has an extension to a
linear ordering $<_0$ of the events. 
\begin{claim} Properties FS0, FS1, and FS2 hold, and hence $<_0$ is a linearization of the $\precc$
relation which proves Theorem \ref{MT}. 
\end{claim}
Proof of Claim.  Note first that if $a<_0 b$ are \Add/\Rem\ events then $a\precc b$ because otherwise we would have $b\precc a$ (by the comparability property in A0), and then $b<_0 a$ would be in contradiction to $a<_0 b$.

Also note that every event has only a finite number of predecessors in $<_0$  (because in the $<$ ordering every event has only a finite
number of events of which it is not a predecessor; see (\ref{LFT})).
 
\begin{description}

\item[Proof of FS1:] Let $a$ be an event such that $\Op^1(a)$ holds. If $\Add(a)$ or $\Rem(a)$, then by Lemma \ref{L1} $\gamma(a)\precc a$
and hence $\gamma(a)<_0 a$ follows (since $<\subseteq <_0$). If $\Cnt^1(a)$, then $\gamma(a)\Rightarrow a$ by (\ref{E5}), and hence $\gamma(a)<_0 a$ again.
The second line of (\ref{E1}) is obvious and follows from (\ref{E4}) in case $\Add^1(a)\vee\Rem^1(a)$, because the $<$ and $<_0$ orderings agree on the \Add/\Rem\
events which are linearly ordered. To prove the second line of (\ref{E1}) when $a$ is a $\Cnt^1$ event
consider any $\Rem^1$ event $r$ such that $\gamma(r)=\gamma(a)$. By equation (\ref{E6}) in the definition of $\Rightarrow$,
$a\Rightarrow r$, and hence no $r$ can contradict (\ref{E1}).

\item[Proof of FS2:] Suppose that $\Op^0(b)$ and $a<_0 b$ is such that $\Add^0(a)\wedge \kval(a) =\kval(b)$. We have to prove that

\begin{equation}
\label{E9}
 \exists c (a<_0 c <_0 b \wedge \Rem^1(c)\wedge a  = \gamma(c))
\end{equation}
\end{description}
Suppose first that $b$ is some $\Add^0$ or $\Rem^0$ event. Then $a<_0 b$ implies that $a<b$.
By A2 there is an event $s$
such that 
\begin{equation}
\label{E10}
\Rem^1(s)\wedge a=\gamma(s)\wedge s<\End(b)).
\end{equation}
 Since $\End(b)=b$, in this case, $s<b$ follow and hence $s<_0 b$,
 and this proves (\ref{E9}) in case $b$ is some $\Add$ or $\Rem$ event.

Now suppose that $b$ is a $\Cnt^0$ event. Then $a<_0 b$ does not immediately imply that $a<b$. But if $a\not<b$, then
(\ref{E8}) of Definition \ref{DA} applies and $b\Rightarrow a$ follows, which implies that $b<_0 a$ in contradiction to the assumption $a<_0 b$.
Thus, $a< b$ even in case $\Cnt^0(b)$, and hence A2 applies and there exists an event $s$ such that (\ref{E10}) holds.
Then item (\ref{E7}) applies, and $s\Rightarrow b$ follows. Thus $s<_0b$ as required. This ends the proof of Theorem \ref{MT}.

\section{The Simpler Lazy Set algorithm}   
\label{S4}
Equipped with Theorem \ref{MT}, it remains to present the Lazy Set algorithm and to prove that  all its executions satisfy
the Lazy Set axioms of Figure \ref{NL1} (which will readily entails linearizability). But the problem (which explains why we are not yet
ready to finish the proof) is that in order to follow this scheme we must present executions of the algorithm as Tarskian system execution.
Usually executions are presented as history sequences of states (also called {\em runs}, see for example \cite{LynchTuttle}),
but it is meaningless to say that a history satisfies a certain sentence in $L_{LS}$ (for it is not a temporal
logic sentence). It is necessary to convert a history sequence into a Tarskian structure in order to apply Theorem \ref{MT}.
 Generally speaking, the way to do it is to take the steps of the history as the events on which predicates and functions are defined and thus to define a Tarskian structure.
But a careful exposition of this conversion is not completely trivial and its main ideas may be lost in the forest of its details.  
So in this section we consider the Simpler Lazy Set algorithm (see Figure \ref{DP1})  and present its linearizability in what we 
hope to be a clearer exposition. 

We assume a fixed, infinite set $A$ of addresses, with two pre-assigned addresses \HH\ and \T. Every address $a\in A$ contains 
two fields: $a.val$
in $\Nat\cup \{-1,\infty\}$
and $a.\next$ in $A$. We have fixed value for the head and tail addresses, $\HH.val = -1$ and $\T.val=\infty$. For any
 $a\in A\setminus \{\HH,\T\}$, the type of $a.val$ is  $\mathbb N$.

The Simpler Lazy Set algorithm is governed by the code at line $0$: process $p$ repeats forever executions of the \AD\ and \RM\ 
actions and the \CONTAINS\ procedure (with natural number parameters). 

\begin{figure}[ht]

\framebox[110mm]{\hspace{-40mm}
\begin{minipage}[t]{100mm} 
\begin{tabbing}
****\=***\=**********\=********\=\kill

\>The Simpler Set algorithm of process $p$:\\ 
 
\> 0. \> Pick some $x\in \Nat$ and \goto\ one of 1, 2, 3.1 \\ 
\end{tabbing}
\end{minipage}
}

\framebox[110mm]{\hspace{-43mm}
\begin{minipage}[t]{100mm} 
\begin{tabbing}
****\=***\=**********\=********\=\kill
 
   
\>1.\>  $\s:= \AD(x)$; \return\, $\s$; \goto\ 0;\\ 
\>       -----------------------------------------------\\
\>2.\> $\s:= \RM(x)$; \return\, $\s$; \goto\ 0;\\
\end{tabbing}
\end{minipage}
}
\framebox[110mm]{ \hspace{-40mm}
\begin{minipage}[t]{100mm} 
\begin{tabbing}
****\=***\=******\=******\=\kill
\> $\CONTAINS(x)$  \\

\>  3.1. \>\  $\curr :=\head$; \\

\> 3.2. \>  \  \Repeat \\              

\> 3.3.\> \   $\curr:=\curr.\next$;\\

\> 3.4. \>\ \until\ $\curr.\kval\geq x$;\\               
 
\> 3.5. \> \ \IF\   $curr.\kval =x$ \THEN\ \return\ $\s=1$\\
\>\> \  \ELSE\ \return\ $\s=0$. \goto\ $0$.

\end{tabbing}
\end{minipage}
}

\caption{The Simpler Lazy Set algorithm. $\AD(x)$ and $\RM(x)$ for $x\in \mathbb N$ are actions, and the $\s$ value
they return is in $\{0,1,f\}$.}
\label{DP1}
\end{figure}
 
In order to explain the working of the Simpler Lazy Set algorithm, we have to define its states, steps, and histories. First, however,
we note the points for which the Simpler algorithm is simpler . 
The \CONTAINS\ protocol of the Simpler algorithm is exactly the same as the \CONTAINS\ of the full algorithm (see Figure
\ref{FEX1}). The simplification of the Simpler algorithm is in the \AD\ and \RM\ protocols which are a single action each. 
The full Lazy Set algorithm has its  \ADD\ and \REMOVE\ protocols which first find addresses \pred\ and \curr\ that are (hopefully) on the main branch, and then determines the new values of the \next\ fields. In order to avoid any interferences  from other processes, the fuller protocols use locking of addresses, a feature not needed here.
Moreover, the fuller \ADD\ and \REMOVE\ protocol uses 
 an additional field in its addresses, called {\em marked},  which is not needed in the Simpler algorithm.  This usage of locks and of the {\em marked} field necessarily complicates the Lazy Set  algorithm and can lead to the status of the operation being $f$.
The Simpler algorithm is completely impractical, but it retains the main difficulty of the full algorithm and hence its value.
 That is,  a correctness proof of the Simpler algorithm that relies on linearization points must find them out of the \CONTAINS\ executions. Instead of using linearization points, however, we are going to prove 
that every execution of the Simpler Lazy Set algorithm satisfies the properties of Figure \ref{NL1}. 
(The predicates \Add, \Rem, and \Cnt\ that appear in that figure represent \AD, \RM, and \CONTAINS\ events.)   
This will imply linearization by the main Theorem (\ref{MT}). 

 Usually, a state is represented as a function that assigns values to the state variables, but we prefer to define
states as structures, because this leads to a clearer description of the state situation. As any structure, a state structure
is an interpretation of a certain language. A state structure of the Simpler Lazy Set algorithm is an interpretation of the
$\SimplerLSstate$ which we define now.  
\begin{definition}
\label{SimplerLang}
The  $\SimplerLSstate$ language is a multisorted predicate language with the following symbols.
\end{definition}
The language contains two sorts: \Address\ and \Number. They have fixed interpretations that do not depend on the interpreting
structure; \Address\ is some countably infinite set, $A$, and $\Number = \Nat\cup\{-1,\infty\}$. There is a binary relation symbol $<$
(for the ordering on \Number), a unary predicate, \PActive, over the \Address\ sort, and two function symbols (that correspond to the
\next\ and \kval\ fields): 
\[ \PNext:\Address\setminus \{\T\} \to \Address, \text{and } \Val:\Address \to \Number.\] Then there is a list of names called {\em constants}.
The intention is to make the resulting states appropriate for the description of states of the Simpler Lazy Set algorithm. So {\em constants} contains the following names.
\begin{enumerate}
\item The program counters $PC_p$ for every process $p$. The possible values of $PC_p$ are the line numbers of the Simpler
algorithm: $0,1,2$ and $3.1,\ldots,3.5$.
\item $status_p$ with values in $\{0,1,f\}$, and $\curr_p$ with values that are addresses.  The parameters $x_p$ are with values in \Nat. 
\item \HH\ and \T\ are two fixed addresses.
\end{enumerate}
It is convenient to have $\bot$ as an ``undefined'' value of  $\SimplerLSstate$.

\begin{definition}
\label{Def4.1}
A state structure $S$ of the Simpler Lazy Set algorithm is an interpretation of the $\SimplerLSstate$ language. $S$ has the form
\end{definition}
 \[ S=(A,\Number; \PActive^S,\PNext^S,\Val^S;constants)\] where $A$ is the \Address\ sort of $S$, a countably infinite set of 
addresses (also called nodes),
 $\Number = {\mathbb N}\cup \{-1,\infty\}$. The set $A\cup\Number$ is the {\em universe} of $S$, it is a fixed set for all
possible state structures.  $\PActive^S\subset A$ is the set of active addresses,
 $\PNext^S:\PActive^S\setminus\{\T\}\to \PActive^S$, 
 and $\Val^S:\PActive^S \to \Number$ are functions, and {\em constants} is a list of names where, for every name $c$, $c^S$ is the interpretation
 of $c$ as a member of the universe of $S$. 

The idea is that given any state $S$ and address $a\in A$, $\PNext^S(a)$
is the value of field $a.next$ in $S$ and $\Val(a)$ is the value of field $a.val$ in $S$.

A {\em path} in state $S$ is a sequence of active
 addresses $a_1,\ldots,a_m$ (where $m>1$) such that for $i<n$ $\PNext^S(a_i)=a_{i=1}$.
We say that such a path leads from $a_1$ to $a_m$. Each $a_i$ is said to be on the path. 

\begin{definition}[Normal states of $\SimplerLSstate$]
\label{DefNS1}
We say that state structure $S$ that interprets $\SimplerLSstate$ is {\em normal} if the following two conditions hold.
\end{definition}
\begin{enumerate}
\item $A$ is infinite and fixed (as we have said), but the set of active addresses is finite. Addresses \HH\ and \T\ are active and with values $-1$ and $\infty$ respectively. If $a\in \PActive \setminus \{\HH,\T\}$ then $\Val(a)\in\mathbb N$.

\item If $a\neq\T$ is an active address and $b=\PNext^S(a)$, then $b$ is an active address and
$\Val(a)<\Val(b)$. 
 
\end{enumerate}
If $S$ is a normal state and $a$ is any active address, then the path $a,\PNext^S(a),\ldots$ that starts with $a$ and is defined by the rule \[\IF\ a_i\neq \T\ \THEN \ a_{i+1}=\PNext(a_i)\]
is a sequence of active addresses with increasing values. Since there is only a finite number of
active addresses in $S$, every such sequence is finite, and this can only be because its last
address is \T\ (over which \PNext\ is not defined).  Hence any active address $a$ is on a unique path
from $a$ to \T. In particular the
This  shows that there are no loops in the path relation, and that 
 any maximal path is finite and ends with  \T\ as its last node. The path from \HH\ to \T\ is said to be the main branch of $S$.
\begin{definition}
\label{Dremoved}
 If address $a$ is active in a normal state $S$, but $a$ is not on its main branch, then we say that
$a$ is {\em physically removed} (or just {\em removed}). 
If $a$ is not on the main branch then the path from $a$ to \T\ is said to be {\em secondary}.
\end{definition}

The initial state structure 
\[ S_0 =(A,{\mathbb N},\{-1,\infty\}; \PActive^0,\PNext^0,\Val^0, constants)\]
is defined by setting $\PActive^0=\{\HH,\T\}$, $\PNext^0(\HH)=\T$, and $\PNext^0(a)=\bot$ for any other address $a$.
For the program counters, $PC_p$, they are all set to line 0 in their code (ready to start $\AD(x)$, $\RM(x)$ or $\CONTAINS(x)$), and $\Val(a)=\bot$
for every $a\in A \setminus \PActive^0$. Clearly the initial state is normal.

 We define now actions
on these state structures. The main actions are \RM(x) and $\AD(x)$, which take a parameter $x\in \mathbb N$, and the assignment actions
such as executions of $b:=a.\next$ where $a$ and $b$ are address variables. 
 These actions are defined by means of ``execution triples''.

\begin{definition}
\label{D4.4} Execution triples (also called {\em steps}) are of the form $(S,e,T)$ where $S$ and $T$
are state structures and $e$ is an action on which a status value  $\chi(e)\in \{0,1,f\}$, and an address $\adr(e)$ are defined.
\end{definition}
\begin{enumerate}
\item There are three types of invocation actions: \AD, \RM, and \CONTAINS. An invocation by process $p$
is enabled on any normal state structure $S$ when $PC_p^S=0$. Then $(S,e,T)$ is an invocation of $\AD(x)$
when $x_p^T\in\Nat$ and $PC_p^T=1$ (and no other variable changes). A response to this invocation is an 
$\AD(x)$ action defined in item 2 below. 

A triple $(S,e,T)$ is a  $\RM(x)$ invocation when $PC^S=0$, $x_p^T\in\Nat$ and $PC_p^T=2$. Responses 
are $\RM(x)$ actions defined in item 3 below. Any $\AD(x)$ or $\RM(x)$ actions may fail.
A triple $(S,e,T)$ is a failed action when $PC_p^S=1,2$, $PC_p^T=0$, and $\chi(e)=f$. 
In any other aspect $T$ and $S$ are equal. (The necessity of failures appears with the fuller Lazy Set algorithm of Section \ref{SLSA}. The reader may ignore this possibility for the moment.) 
 
$\CONTAINS(x)$ executions are high level events consisting of a sequence of reading actions.
 In an invocation of $\CONTAINS(x)$, control variable $PC_p$
changes from $0$ to $3.1$, and the response is the return action in which $PC_p$ changes from $3.5$ back to $0$. The status value of a \CONTAINS\ execution is never $f$.

\item 
An $\AD(x)$ action by process $p$ is enabled on any normal state structure $S$ when $PC_p^S=1$. An $\AD(x)$ execution triple has the 
form $(S,e,T)$ where $T$ is the state structure defined by the following procedure. $PC_p^T=0$ (making the system ready for
the next invocation). 
Let $\p$  be the address on the main branch of $S$ such that $\Val^S(\p)<x$ but $\Val^S(\PNext^S(\p))\geq x$.
There are two cases: 
\begin{enumerate}
\item If $\Val^S(\PNext^S(\p))= x$ then $T=S$ and we set $\chi(e)=1$ and $\adr(e)=\PNext^S(\p)$ (the address of value $x$ that
was found on the main branch). 

\item If $\Val^S(\PNext^S(\p))> x$, then surely there is no address of value $x$ on the main branch of the normal structure $S$.
In this case
choose
$a\in A\setminus \PActive^S$ (said to be a ``new'' address). Then $\PActive^T=\PActive^S\cup\{a\}$, $\Val^T(a)=x$, and
$\PNext^T( \p)=a$ and $\PNext^T(a)=\PNext^S(\p)$. We set $\chi(e)=0$ and $\adr(e)=a$ in this case. Clearly, $T$ is normal 
state with $a$ on its main branch.
We say in this case that $e$ {\em activates} $a$.
\end{enumerate}
In both cases we say that $e$ is an \AD\ action of value $x$ and its status is $\chi(e)$.
In both cases $\adr(e)$ is the address of value $x$ found on the main branch or added to the main branch by $e$
 (when $\chi(e)=1$ or $\chi(e)=0$).

\item A $\RM(x)$ action by process $p$ is enabled on any normal state structure $S$ when $PC_p^S=2$. Then $(S,e,T)$ is a $\RM(x)$ triple 
when $PC_p^T=0$ and the following holds.
There are two cases. 
\begin{enumerate}
\item In the first case there is on the main branch of $S$ an address, denoted $\cu$, of value $x$ (since $x\in\Nat$, $\cu\not\in \{\HH, \T\}$). 
In this case, a $\RM(x)$ execution triple has the form $(S,e,T)$ where $T$ is defined by the following procedure.
Let $\pred$  be that node on the main branch of $S$ such that $\PNext^S(\pred)=\cu$ (since $x\in \Nat$ there is such a node, and
 it is possible that $\pred=\HH$). Then $\PNext^T(pred)=\PNext^S(\cu)$. Clearly $T$ is also normal, and address \cu\ is no longer on the 
main branch of $T$. In fact the main branch of $T$ is equal to the main branch of $S$ without \cu. We say that $e$ is a \RM\ action of value $x$ and we set $\chi(e)=1$ and $\adr(e)=cu$. We
also say that address $\cu$ is {\em physically removed from the main branch of} $T$.   

\item There is no address on the main branch of value $x$. In this case  $T=S$ (except for the $PC_p$ value of course) and $e$ is a $\RM$ action of value $x$ and
with status $\chi(e)=0$. (The \adr\ function is not defined on $e$, i.e. $\adr(e)=\bot$.)
\end{enumerate}

\item For any address variables $a$ and $b$, an assignment (reading) action such as $b:=a.\next$ is enabled for process $p$ in a normal state $S$ when $PC_p$ points to that instruction and address $a$ is active  (and different from $\T$).  
Then  $(S,e,T)$ is an execution triple of action $b:=a.\next$ when $PC_p^T$ is the next instruction, $b^T=\PNext^S(a)$ and
 $T$ is equal to $S$ as far as the \PNext\ and \Val\ functions go. Assignment instructions appear in the $\CONTAINS(x)$ protocol.

\end{enumerate}
Note that if $(S,e,T)$ is a triple then $S$ and $T$ are both normal states.

A {\em history sequence} of the Simpler Set algorithm is an alternating sequence of states and actions 
 \[ H= (S_0,e_0,S_1,\ldots,\ldots,S_j,e_j,S_{j+1},\ldots)\] (usually infinite) such 
that $S_0$ is the initial state structure, and for every $j$ $(S_j,e_j,S_{j+1})$ is an execution triple, as defined above, by one of the processes.  So $e_j$ can be some invocation, some \AD\ action (which is taken to be a response), some \RM\ action, or some line $3.i$ execution (for $i=1,\ldots,5$). Note that all states in the history are normal.

The {\em life cycle} of every address $a$ (different from \HH\ and \T)  starts as an inactive address, and then it may become
active by some $\AD(x)$ action $e$ in the history with $\chi(e)=0$. This action $e$ activates $a$ with  value $x$ and introduces $a$ to the main
branch into its position according to its value. Note that the value of $a$ (that is $x$) never changes afterward and $a$ 
stays active forever. Then, it is possible that
$a$ is removed from the main branch by some $\RM(x)$ action (but it remains active). Even after removal, there is a path from $a$ to \T\ 
(as all states in the history are normal). An active address that is not on the main branch of a state $S$ is said to be ``physically removed'' or just ``removed''.

\begin{lemma}
\label{lemma4.6} 
Let $(S,e,T)$ be some \RM\ action that removes address $d$ from the main branch of $S$. Let $c$ be that
address on the main branch of $S$ such that $\PNext^S(c)=d$ ($c$ exists because $d\neq \HH$). Say $d'=\PNext^S(d)$. Let $a$ be any active address
in $S$ (on its main branch or out of that branch); let $Q$ be the path from $a$ to $\T$ in $S$, and $R$ be the  
path from $a$ to \T\ in $T$. Then either $R=Q$ or else $R=Q\setminus \{d\}$. 
In fact there are two cases in the proof.
\begin{enumerate}
\item
If $c$ is not in $Q$ then $Q=R$ (even when $d$ is in $Q$). 
\item
If $c$ is in $Q$ then $d$ and $d'$ are also in $Q$, and in this case $R=Q\setminus \{d\}$.
\end{enumerate} 
\end{lemma}
Proof By definition of
the \RM\ triples, the only difference between $\PNext^S$ and $\PNext^T$ is that $\PNext^T(c)=d'$ (whereas
$\PNext^S(c)=d$). The proof follows directly by comparing $R$ with $Q$ in the two cases. In case $c$ is not
on $Q$, for every $x\in Q$, $\PNext^S(x)=\PNext^T(x)$, and hence $Q$ is also a path in state $T$. So
$Q=R$ is this case. 

If $c\in Q$ then also $d=\PNext^S(d)$ is in $D$ since $D$ is a path and $d\neq \T$. So $d'=\PNext(d)$
is also in $Q$ (but $d'=\T$ is a possibility). Anyhow, $Q=P\setminus \{ d\}$ follows since $\PNext^T(c)=d'$ is the only difference between the two states.  \qed

The following lemma implies that if an address is physically removed at state $S_i$ in a history then it remains removed at any later state.
\begin{lemma}
\label{Lem4.5}
For every action triple $(S,e,T)$ and address $a$, if $a$ is removed at state $S$ then it is still removed at $T$.
\end{lemma}
Proof. Recall (Definition \ref{Dremoved}) that an address $a$ is said to be removed at a state if it is active there but not on
the main branch. We thus have to check that neither \AD\ actions nor \RM\ actions can introduce into main branch an active address that is not already on that
branch. The only action that can add an address to the main branch is an \AD\ action (It is easy to check that no \RM\ action adds an address to the main branch.) But an \ADD\ triple
$(S,e,T)$ activates an address $a'$ to the main branch only in case $a'$ is not active in $S$
(by definition of activation triples). Hence if $a$ is removed at $S$ it remains removed at $T$. \qed

The following definition relies on the observation that if some address $a$ that is different from \HH\ and \T\ is active in
state $S_j$ then there is $i<j$ such that $e_i$ is an activation of $a$. For the proof, note that in the initial state $S_0$
only \HH\ and \T\ are active, and hence if $a$ is active in $S_j$ there exists some index $i<j$ such that $a$ is inactive 
in $S_i$ but active in $S_{i+1}$. Checking the different types of triples, we immediately realize that $(S_i,e_i,S_{i+1})$
has to be some $\AD(x)$ triple. Moreover, since the value $\Val(a)$ of an address never changes after activation, $x$ has to
be the value of address $a$ in $S_j$.
An active address can have just one activation, and hence we can define a function as follows.
\begin{definition}
\label{Defact}
Consider an arbitrary state $S_j$ in $H$; if address $a$, different from \HH\ and \T, is active and with value $x$ in $S_j$, 
then there exists some activation event $e_i$ for $i<j$ that activates $a$. So $\AD(e_i)$, $\Val(e_i)=x$, $\chi(e_i)=0$, and
$\adr(e_i)=a$. We write
\begin{equation}
e_i=\activation(a). 
\end{equation}
We note that the definition of $\activation(a)$ does not depend on $S_j$.
\end{definition}

It is convenient to look at any history sequence $H$ as a Tarskian structure, or preferably to transform $H$ into a 
Tarskian structure $M_H$ defined as follows. The universe
of $M_H$ consists of the low level actions $\{e_i\mid i\in\Nat\}$, the higher level \CONTAINS\ executions that will be defined shortly, and the set $A$ of addresses. 
Predicates and functions are defined over these actions as follows. 
\begin{enumerate}
\item Every process name $p$ is taken to be a predicate on the actions. So $p(e)$ says that $e$ is an event by process $p$.

\item Predicates \AD\ and \RM\ are defined over the events of $M_H$: \AD$(e_i)$ iff 
$(S_i,e_i,S_{i+1})$ is an
\AD\ action, and likewise predicate \RM\ is defined\footnote{So \AD, for example, is an overloaded
symbol: \AD\ as in $\AD(x)$ that appears in the code is not a predicate, and in formula $\AD(e)$
$e$ is not a number variable.}. The $\chi$ and \Val\ functions are defined over
the \AD\ and \RM\ actions. For example, if the triple $(S_i,e_i,S_{i+1})$ is an execution of $\AD(x)$ with status $0$,
then we define $\AD^0(e_i)$, $\chi(e_i)=0$, $\Val(e_i)=x$, and $p(e_i)$.

If $e$ is an \AD\ action  or an $\RM^1$ action, then $\adr(e)$ is defined by Definition \ref{D4.4}.

\item Recall that for every state $S$ we have the function $\PNext^S$ and $\Val^S$ defined over the set of addresses $A$, and the predicate
$\PActive^S\subset A$. Instead of introducing the states $S_i$ as objects of the structure $M_H$ (which is a reasonable possibility), 
we define functions and predicate
 of two variables as follows.
\begin{enumerate}
\item For addresses $a,b$ and $x\in\Nat$,
 $\PNext(a,e_i)=b$ holds in $M_H$ when $\PNext(a)=b$ holds in $S_i$. $\Val(a,e_i)=x$ holds in $M_H$ when $\Val(a)=x$
holds in $S_i$. 
\item $\PActive(a,e_i)$ holds in $M_H$ when $\PActive(a)$ holds in $S_i$.
\end{enumerate}
\item For any address $a$ we define $\PActive(a)$ in $M_H$ iff for some $i$ $\PActive(a,e_i)$ holds. In this case, $\PActive(a,e_j)$
holds for every $j\geq i$. If $\PActive(a)$ then we define $\activation(a)$ as in Definition \ref{Defact}.
\end{enumerate}

We noted that if address $a$ is active in some state $S_i$ in the history then it remains active in every later state.
Another way to express this is \[\PActive(a,e_i)\Rightarrow \forall e_j(e_i< e_j\Rightarrow \PActive(a,e_j)).\]
It follows that no address can be activated twice. The reason is that if $(S,e,T)$ is an activation of some address $a$ (see item
2 in Definition  \ref{D4.4}) then $a$ is not active in $S$ (it is a ``new'' address).

In addition to actions, we introduce to $M_H$  higher level \CONTAINS\ operation executions.
Some actions in $M_H$ belong to executions by the different processes of instructions of the 
 \CONTAINS\ protocol. These actions correspond to lines $3.1,\ldots,3.5$ of the \CONTAINS\ protocol, and are 
assembled into higher level operation executions. An operation execution $E$ by process $p$ of the $\CONTAINS(x)$ protocol 
is a sequence of actions that represent an execution of that protocol. If $E$ is a $\CONTAINS(x)$ operation execution, then
$\Begin(E)$ is the first action in $E$ (the action that executes $\curr:=\HH$ at line 3.1), and $\End(E)$ is the last action in $E$ (the return action). 
It is convenient to define $\Begin$ and $\End$ on actions as well. For every \AD\ and \RM\ action $e$ we define
 $\Begin(e)=\End(e)=e$.

A temporal precedence relation $<$ is defined for $M_H$.
 The (lower level) actions in $H$ are linearly ordered: $e_i<e_j$ iff $i<j$, but the ordering on all the events (actions and operation executions) is a partial ordering: for events $E_1,E_2$ (actions or operation executions) we define $E_1<E_2$ 
 iff $\End(E_1)<\Begin(E_2)$. 

The function \activation\ defined above in \ref{Defact} is considered  as a function of $M_H$.

 This ends the definition of the structure $M_H$.

Our aim is to prove the following theorem which implies the linearizability of the Simpler algorithm (by Theorem \ref{MT}).
\begin{theorem}
\label{MSimpler}
If $H$ is a history sequence of some execution of the Simpler Lazy Set algorithm, then the structure $M_H$ satisfies the Lazy Set axioms
of Figure \ref{NL1}.
\end{theorem}
Proof.
We begin with a detailed analysis of an arbitrary $\CONTAINS(x)$ operation $E$ by some process $P$. Consider, for simplicity of expression, only
the actions in $E$ that determine the successive values of variable \curr\ of $P$ and enumerate them in their execution order:
$\{e_{j_0},\ldots,e_{j_m}\}$.  Let $S_{j_0},\ldots,S_{j_m}$ be the corresponding states of $H$. That is,
$(S_{j_i},e_{j_i},S_{j_i+1})$ is
the triple in $H$ that characterizes the action $e_{j_i}$. Let $a_0,\ldots,a_m$ be the corresponding addresses that
variable \curr\ takes. I.e. $a_i= curr^{S_{j_i} }$. Thus $a_0=\HH$, and for $0<i\leq m$, 
\begin{equation}
\label{Eq17}
a_i=\PNext^{S_{j_i} }(a_{i-1}).
\end{equation}
Each address $a_i$ is active in $S_{j_i}$ (and in subsequent states) because these states are normal, and by the observation
that an active address stays active forever in a history. 
The sequence of addresses $X=\{a_{0},\ldots,e_{m}\}$ 
 is not necessarily a path in any of the structures $S_i$ of the history.
We say that $X$ is the {\em pseudo-path} of $E$. Even though $X$ is not a path it is still true that every $a_i$ is active and that $\Val(a_{i-1})<\Val(a_{{i}})$.
(This follows immediately from equation (\ref{Eq17}).)
The corresponding sequence of states $S_{j_0},\ldots,S_{j_m}$ is called the {\em state sequence of} $E$.

The \until\ condition at instruction 3.4 implies that, for $i<m$, $\Val(a_i)<x$ (we write $\Val(a_i)$ rather than $\Val^{S_i}(a_i)$ since the value of an active address
never changes). Thus 
\begin{equation}
\label{EQbe}
\Val(a_0)<\Val(a_1)<\cdots < \Val(a_{m-1})<x,
\end{equation}
 and hence $E$ must be finite.
 For the last index $m$ we have that $a_m =\PNext^{S_{j_m}}(a_{m-1})$
and 
\begin{equation}
\label{EQ19}
\Val(a_{m-1})<x\leq \Val(a_m).
\end{equation}
 In case $\Val(a_m)=x$ the status $\chi(E)$ returned by $E$ is $1$, and in case $\Val(a_m)>x$
the status is $0$. We note that the status of $E$ is $1$ even in case $a_m$ is a removed address, that is an active address not on the main branch of $S_{j_m}$. 

Suppose that $b\neq \HH$ is an address on the main branch of $S_i$ but is no longer on the
main branch of $S_{i+1}$. Then it must be the case that $(S_i,e_i,S_{i+1})$ is a \RM\
action that removes $b$ from the main branch (i.e. $b=\adr(e_i)$). 
 The following lemma generalizes this observation. 

\begin{lemma}
\label{Lem4.8}
Let $S_i$ be a state in history $H$ and $a\neq b$ two active addresses  in $S_i$ such that there is 
a path $Q$ from $a$ to $b$ in $S_i$. Suppose that in $S_{i+1}$ there is no path from $a$ to $b$ 
(equivalently,
$b$ is no longer on the path from $a$ to \T\ in $S_{i+1}$.) 
 Let $b^-$ be that address on $Q$ that precedes $b$ (i.e. such that
$\PNext^{S_i}(b^-)=b$). Then the following hold:

\begin{enumerate}
\item $b^-$ and $b$ are on the main branch of $S_i$, and
\item $(S_i,e_i,S_{i+1})$ is a \RM\
triple that removes $b$ from the main branch by setting \[\PNext^{S_{i+1}}(b^-)=\PNext^{S_i}(b).\] 
\end{enumerate}

\end{lemma}
Proof. 
What could step $(S_i,e_i,S_{i+1})$ be? It cannot be an assignment triple (which do
not change the tree structure). It is not an \AD\ step, since an \AD\ step may make the path from $a$ to $b$ longer (by adding an address) but
cannot destroy an existing path. So $(S_j,e_j,S_{j+1})$ is necessarily a $\RM(x)$ triple that removes  some address $d$ of value $x$ and 
 which lies on the main branch of $S_i$ (as the definition of these triples requires).
If $d^-$ is that address on the main branch of $S_i$ such that $\PNext^{S_i}(d^-)=d$, then
$\PNext^{S_{i+1}}(d^-)=\PNext^{S_i}(d)$. 

We claim that $d^-=b^-$. Once we prove this claim, it follows that $d=b$ and the proof is concluded.
We first note that $d^-$ is on $Q$. If this were not so then the path $Q$ would remain a path from 
$a$ to $b$ in $S_{i+1}$ as well. Moreover $d^-=b^-$ because otherwise there are only two possibilities:
 (1) that
$d^-$ is earlier than $b^-$ on $Q$, and then there would be an even shorter path from $a$ to $b$ in 
$S_{i+1}$, and (2) that $d^-=b$ in which case $Q$ remains a path from $a$ to $b$ in $S_{i+1}$.
  \qed

\begin{cor}
Let $S_i$ be a state in history $H$ and $a\neq b$ two active addresses  in $S_i$ such that there is 
a path  from $a$ to $b$ in $S_i$, but for some $k>i$ there is no path from $a$ to $b$ in $S_k$.
If $j$, is the maximal index such that  $j\leq k$ and there is a path from $a$ to $b$ in $S_j$,
then $i\leq j<k$ and $(S_j,e_j,S_{j+1})$ is a   

\end{cor}

\begin{lemma}
\label{L4.9}
Suppose that in state $S_i$ of history $H$ two active addresses $a$ and $b$ are such that $a\neq \T$ and the path from $a$ to \T\ does not contain $b$. Then also in
$S_{i+1}$ the path from $a$ to $\T$ does not contain $b$.

Hence if $b$ is not on
the path of $S_i$ that leads from $a$ to $\T$. Then for every $j>i$ $b$ is not on the path in $S_j$ that leads from $a$ to \T.
\end{lemma}
Proof.
 Consider the two possibilities for the triple $(S_i,e_i,S_{i+1})$.
If $e_i$ is an \AD\ action then for active addresses $a,b$ in $S_i$ there is a path from $a$ to $b$ in $S_i$ iff there is one
in $S_{i+1}$.

Suppose now that  $e_i$ is some \RM\ action that removes address $d$ from the main branch of $S_i$, and let $c$ be that address on
the main branch such that $\PNext^{S_i}(c)=d$. 
 Let $Q$ be the path from $a$ to \T\ in $S_i$, and $R$ be the path from $a$ to \T\ in $S_{i+1}$.
By lemma  \ref{lemma4.6} there are two possibilities. If $c$ is not on $Q$ then $R=Q$ and the lemma is obvious in this case.
But if $c$ is on $Q$ then $R$ is a proper subset of $Q$, and hence surely $b$ is not in $R$. \qed

The function $\gamma$ is defined on the $\Op^1$ events  ($\AD^1$, $\RM^1$, and $\CONTAINS^1$ events) and returns
$\AD^0$ actions, and is defined as follows.

\begin{definition}
\label{DefGamma}
\begin{enumerate}
\item
If $e$ is some $\AD^1$ or $\RM^1$ event and $a=\adr(e)$ then we define $\gamma(e)=\activation(a)$.
($\adr$ is defined in \ref{D4.4}, and \activation\ in \ref{Defact}.)
\item If $A$ is some $\CONTAINS(x)$ operation execution with $\chi(A)=1$ and $X=\{a_{0},\ldots,a_{m}\}$ is the pseudo-path of
$A$, then address $a_m$ is active and of value $x$. Define $\gamma(A)=\activation(a)$. Then $\gamma(A)$ is an $\AD^0(x)$ event $e$ such that
$e<e_{j_m}$ (and hence $e<\End(A)$).

\end{enumerate}
\end{definition}

We are now ready for the proof of Theorem \ref{MSimpler}. Property A0 is obvious, and we concentrate on the the two remaining
properties A1 and A2, reworded as follows for the Lazy Set protocols.

A1: 
For every $\Op^1$ event $A$:
\begin{equation}
\label{E4n}
\begin{aligned} 
&\AD^0(\gamma(A))\wedge \kval(\gamma(A)) =\kval(A)\wedge \gamma(A) <\End(A) \wedge\\
& 
\neg \exists R ((\RM^1(R)\wedge \gamma(R) = \gamma(A) \wedge \gamma(A)\precc R \precc A)
\end{aligned}
\end{equation}

A2:
For every events $A$ and $B$:
\begin{equation}
\label{E4na}
\begin{aligned}
&if  \ \Op^0(B) \wedge  \AD^0(A)\wedge  A\precc B \wedge \kval(A)=\kval(B), \\ 
&then \ \exists R (\RM^1(R)\wedge A=\gamma(R)\wedge R<\End(B)).
\end{aligned}
\end{equation}
 
We Prove A1. The first line of (\ref{E4n}) is a consequence of the definition of $\gamma$ and the observations that were made there. For
the second line, assume towards a contradiction the existence of some $\RM$ action $R$ such that $\chi(R)=1$, $\gamma(R) = \gamma(A)$,
and $\gamma(A)\precc R \precc A$. There are two cases to check which depend on $A$.
\begin{enumerate}
\item For the first case, $A=e_i$ is some $\AD(x)$ or $\RM(x)$ action and 
is such that  $\chi(A)=1$. There is an address $a=\adr(e_i)$  with value $x$ for which we 
 defined $\gamma(e_i)=\activation(a)$. Then $\gamma(e_i)=e_{\ell}$ where $\ell<i$ and $e_\ell$ activates address $a$.
  
We assumed (in contradiction to the second line of (\ref{E4n})) the existence of some 
\RM\ action $R=e_j$ 
such that $\gamma(R)=e_\ell$ and $e_\ell < e_j<e_i$.
Now $\gamma(e_j)=e_\ell$ implies (by definition of $\gamma$) that the address removed by $R$ is the address that $e_\ell$ activates, i.e. $a$.
So $a$ is a removed address in $S_{j+1}$ but is again on main branch of $S_i$. This is impossible by Lemma \ref{Lem4.5}
which says that once removed no address can return to the main branch. \qed

\item For the second case,  $A$ is some $\CONTAINS(x)$ operation execution such that $\chi(A)=1$.
Let
 $X=\{a_{0},\ldots,a_{m}\}$ be the pseudo-path of
$A$, and $S_{j_0},\ldots,S_{j_m}$ be the state sequence of $A$. We noted that every $a_i$ is an active address. Since $\chi(A)=1$,  address $a_m$  has value $x$,
and we defined $\gamma(A)=\activation(a_m)$ as that $\AD(x)$ event that activates address $a_m$. 
 Suppose for a contradiction that there exists some \RM\ action $R=e_j$ such that 
$\gamma(R)=\gamma(A)$ and $\gamma(A)<R <A$.
Thus $R$ removes address $a_m$ and hence $a_m$ is not on the main branch of $S_{j+1}$. 
By Lemma \ref{L4.9}, it is also the case
 that $a_m$ is not on the main branch of $S_{j_0}$, i.e. not on
 the branch that leads from $\HH=a_0$ to \T. We shall prove by induction that
for every index $k$, $0\leq k\leq m$, $a_m$ is not on the path of $S_{j_k}$ that leads from 
$a_{k}$ to \T. 

We noticed that
this is the case for $k=0$. Suppose  for $S_{j_k}$ that $a_m$ is not on its path that leads
 from $a_k$ to \T. By Lemma \ref{L4.9} again, it remains true also in $S{j_{k+1}}$
that $a_m$ is not on the path from $a_k$ to \T. By definition, $a_{k+1}=\PNext^{S_{j_{k+1}}}(a_k)$ and hence surely it is not the case that
$a_m$ is on the path from $a_{k+1}$ to $\T$ in $S_{j_{k+1}}$. 
Finally, in $S_{j_{m}}$, $a_m$ is not on the path from $a_{m}$ to \T,
but this is absurd.

\end{enumerate}

We now prove A2. Suppose that $A$ and $B$ are such that $\Op^0(B) \wedge  \Add^0(A)\wedge  A\precc B \wedge \kval(A)=\kval(B)$
and yet for every $\RM^1$ action $R$ such that $A=\gamma(R)$, $\neg (R<\End(B))$. Since both $R$ and $\End(B)$ are actions,
$\neg (R<\End(B))$ is equivalent to $\End(B)<R$. Let $a$ be the address that $A$ activates (i.e. $a=\adr(A)$), 
and let $x$ be the value of $A$ (which is
the value of address $a$). Say $A=e_i$. There are two cases as above.
\begin{enumerate}
\item 
In the first case, $B$ is some $\AD^0$ or $\RM^0$ actions of value $x$ (since $\kval(A)=\kval(B)$). Say $B=e_j$. So $e_i<e_j$ and
there is no $\RM^1$ action $R$ that removes $a$ and is such that $e_i< R<e_j$. This clearly implies that, for every index $k$ 
such that $i<k\leq j$, $a$ is on the main branch of $S_k$. But since $e_j$ is an $\Op^0$ action, there is no address of value
$x$ on the main branch of $S_j$ (by definition of these actions). A statement contradicted by address $a$.

\item In the second case, $B$ is some $\CONTAINS$ operation execution of value $x$ and with status 
$0$.
Let $X=\{a_0,\ldots,a_m\}$ be the pseudo-path of $B$, and $S_{j_0},\ldots,S_{j_m}$ be its state sequence. Recall equation (\ref{EQ19}) which says that \[\Val(a_{m-1})<x\leq \Val(a_m).\]
 Action $A=e_i$
activates address $a$ and so $a$ is a node of value $x$ on the main branch of $S_{i+1}$. By assumption, if $R$ is any \RM\ action
that removes address $a$  (i.e. such that $A=\gamma(R)$) then $B<R$ (equivalently $e_{j_m}<R$). This implies that $a$ is on the main
branch of any state $S_k$ such that $i<k\leq j_m+1$. We shall prove for every $k$ such that
$0\leq k\leq m$ that 
\begin{equation}
\label{Equ22}
a\ \text{is on the path from } a_{k}\ \text{to} \T\ \text{in } S_{j_k}.
\end{equation}
 Hence 
$\Val(a_{k})\leq\Val(a)=x$, and in particular $\Val(a_{m})\leq  x$. In view of equation (\ref{EQ19}) we get that $\Val(a_m)=x$.
But this implies that the status of $B$ is $1$ in contradiction to our assumption.

We prove (\ref{Equ22}) by induction on $0\leq k\leq m$. For $k=0$, as $a_0=\HH$ and $a$ is on
the main branch of $S_{j_0}$, $a$ is indeed on the path from $a_0$ to \T. Suppose now that $a$ is
on the path from $a_k$ to \T\ in $S_{j_k}$ and $k<m$. Since there is no \RM\ action
between $e_{j_k}$ and $e_{j_{k+1}}$ that removes address $a$, Lemma \ref{Lem4.8} implies that
$a$ is still on the path from $a_k$ to \T\ in $S_{j_{k+1}}$. Recall (equation (\ref{Eq17}))
that $a_{k+1} = \PNext^{S_{j_{k+1}}}(a_k)$. Since $\Val(a_k)<x$ (by equation (\ref{EQbe}))
$a_k\neq a$ and hence $a$ has to be on the path from $a_{k+1}$ to \T\ in $S_{j_{k+1}}$.

\end{enumerate}\qed

\noindent
{\bf A note on stuttering steps.} To the definition of triples (Definition \ref{D4.4}) we can add stuttering triples. We say that
$(S,e,T)$ is a stuttering triple when $S=T$. In this case $e$ is a {\em no-action} event. Stuttering steps were introduced by Lamport
as a convenient way to deal with projections of a system on a smaller one. They are not strictly necessary; an alternative is to fuse
the two states, but this would necessitate a new enumeration of the states in a history. Anyhow, it should be quite obvious that Theorem
 \ref{MSimpler} continues to hold in the presence of stuttering triples. We shall use this remark when proving linearizability of the Lazy Set algorithm. 

\section{The Lazy Set Algorithm}
\label{SLSA}
The Lazy Set Algorithm of \cite{Helleretal} is presented in figures \ref{DP} and \ref{FEX1}\footnote{Actually we have here a variant of
the algorithm of \cite{Helleretal}.}.  It is composed of three protocols (procedures), \ADD, \REMOVE, and \CONTAINS, that are
repeatedly executed by  concurrently operating processes. 
Each process $P$ executes the  generic program of Figure \ref{F6} which repeatedly invokes the three protocols with arbitrary 
parameters $x\in\Nat$. 
 The protocols presented in figures \ref{DP} and \ref{FEX1} are templates that any process can employ by adding its name $P$ as an index to each
of the local variables in the protocol.

\begin{figure}[h!]

\framebox[110mm]{\hspace{-23mm}
\begin{minipage}[t]{120mm} 
\begin{tabbing}
****\=***\=***\=***\=\kill


\> 0 \> \> Pick some $x\in \Nat$ and \goto\ one of\\ 
            
\>  1.1\>  \> (execute \ADD$(x)$); \\

\> 2.1 \> \> (execute $\REMOVE(x)$);\\

\> 3.1 \>\> (execute $\CONTAINS(x)$);\\

\end{tabbing}
\end{minipage}
}

\caption{The generic algorithm of process $P$: repeat forever \ADD, \REMOVE, and \CONTAINS\ executions. The generic code 
consists of a single line: line $0$. Each of the three operations (\ADD, \REMOVE, and \CONTAINS) ends its execution
by returning to line $0$.}
\label{F6}
\end{figure}

\begin{figure}[h!]

\framebox[110mm]{\hspace{-25mm}
\begin{minipage}[t]{120mm} 
\begin{tabbing}
****\=****\=**********\=********\=\kill
\underline{Success  Condition}: \\
\> $SC \equiv \neg \pred.\marked \wedge  \pred.\next = \curr$

\end{tabbing}
\end{minipage}
}

\framebox[110mm]{\hspace{-60mm}
\begin{minipage}[t]{100mm} 
\begin{tabbing}
****\=****\=****\=***\=\kill
$\locate(x)$  \\

 1 \>  $\curr :=\head$; \\

 2 \>    \Repeat \\          

 3.1 \>\> $\pred:= \curr$;\\

3.2 \> \> $\curr:=\curr.\next$;\\

4\> \until\ $\curr.\kval\geq x$;\\     

5 \> \return\ $(\pred,\curr)$. 

\end{tabbing}
\end{minipage}
}

\begin{minipage}[t]{100mm}
\vspace{1mm}
\end{minipage}

\framebox[110mm]{\hspace{-32mm}    
\begin{minipage}[t]{100mm} 
\begin{tabbing}
****\=**************\=*********\=********\=\kill
\ADD$(x)$: \hspace{53mm}\\
1.1 \>$(\pred,\curr) :=\locate(x)$; \\

 1.2\>$\pred.\lock$;\> \\

 1.3 \>\IF\ $\neg SC$ \THEN\\
\> $\pred.\unlock(); \return\ \Failed$; \\
1.4 \>\IF\ $(\curr.\kval=x)$ \THEN \\
 \>$\pred.\unlock()$; $\return\ 1$ ; \\

1.5 \>$\pred.\next := \New(\kval:=x,\next:=\curr)$;\\          
1.6 \>$\pred.\unlock()$; $\return \ 0. $ \\

\>--------------------------------------------------\\

\REMOVE$(x)$: \hspace{53mm} \\
2.1 \>  $(\pred,\curr) := \locate(x)$  ;\\

2.2 \> $\pred.\lock$;\\

2.3 \> \IF\ $\neg SC$ \THEN\\
\> \pred.\unlock(); \return\ \Failed;\\

2.4 \> \IF\ $\curr.\kval > x$ \THEN\\
\>  $\pred.\unlock()$, \return\ $0$;\\

2.5 \> $\curr.\lock$; $d:=\curr.\next$ \\

2.6 \> $\curr.\marked := \TRUE$;\\

2.7\> $\pred.\next := d$;\\

2.8 \> $\curr.\unlock()$; $\pred.\unlock()$; \return\ $1$ 
\end{tabbing}
\end{minipage}
}

\caption{The \ADD\ and \REMOVE\ protocols. The success condition
$SC$ and \locate\ procedure are used by these protocols. 
 Instruction 1.5 of the \ADD\ protocol  is named ``activation'';  Instructions 2.6 and 2.7 
are named ``marking'' and ``physical removal''. }
\label{DP}
\end{figure}

\begin{figure}[ht]

\framebox[110mm]{ \hspace{-30mm}
\begin{minipage}[t]{100mm} 
\begin{tabbing}
****\=***\=***\=***\=\kill
\> $\CONTAINS(x)$  \\

\>  3.1 \>  $\curr :=\head$; \\

\> 3.2 \>    \Repeat \\              

\> 3.3 \> $\curr:=\curr.\next$;\\

\> 3.4 \> \until\ $\curr.\kval\geq x$;\\               
 
\> 3.5 \> \IF\   $\curr.\kval =x$ \THEN\ \return\ $1$ \ELSE\ \return\ $0$.

\end{tabbing}
\end{minipage}
}

\caption{The \CONTAINS\ protocol resembles the \locate\ procedure.} 
 \label{FEX1}
\end{figure}

 Chapter 9 of \cite{HS} contains an excellent
introduction to optimistic lazy set algorithms which gradually introduces algorithms of increasing complexity, explaining the reasons they work in an informal and suggestive language. Our main aim in this paper
is to describe a mathematical framework in which correctness of distributed algorithms can be proved. We too present the Lazy Set algorithm
gradually by dealing with the Simpler Set algorithm firstly. But the role of the Simpler algorithm is not only didactic;
 we shall see that any execution of the full Lazy Set algorithm has a reduct that is an execution of the Simpler algorithm. Thus the correctness of the full algorithm is formally concluded from the already established linearizabilty of the Simpler algorithm. Of course, we have to define the reduct and to prove its properties, but this is much easier than a complete correctness proof.

In the following we clarify some points on the text of the algorithm of figures \ref{F6}, \ref{DP}, and \ref{FEX1}.
We assume a space, \Address, of addresses
 where each address $a$ has three fields: $a.\kval\in \Nat$ (the key value of $a$),
$a.\marked\in \{ \TRUE,\FALSE\}$, and $a.\next \in \Address$. (Note that an address in the space of addresses of the Simpler Lazy Set
algorithm does not have the \marked\ field.)
 Two addresses \head\ and \Tail\ have fixed, pre-assigned values, $-1$ and $\infty$.
We shall define the notion of {\em normal state} and prove that all states that result in an execution of the Lazy Set algorithm are
normal. In a normal state, for any active address $a$ there is a path (of \next\ links) that leads from $a$ to the sink address \T.
The path from \HH\ to \T\ is said to be the main branch and any path from an active node not on the main branch to \T\ is
on a {\em secondary branch}.



In checking whether the success condition, SC, holds,
the process that executes the \ADD\ or \REMOVE\ protocol 
reads the fields  \pred.\marked\ and \pred.\next\  in any order (to some local variables that are
not mentioned in our figures) and then tests condition $SC$. 

All the papers cited in \cite{DD} that deal with the Lazy Set algorithm assume that some unspecified
garbage collection process works in the background. An alternative assumption which we make here is to assume 
an infinite set of ``reserved and inactive addresses'' from which  activations of new addresses are made.
An activation is an execution of $\pred.\next := \New(\kval:=x,\next:=\curr)$ which assigns to the \next\ field
of address \pred\ a new address $a$ that
has never been activated before (and is surely different from the \head\ and \T\ addresses). This activation determines the
key value, $x$, and the \next\ value, \curr, of $a$.
  It is the obligation of the system to ensure that different activations assign different addresses to different executions of
		$\pred.\next$ by
	keeping track of those addresses that are already in \PActive\ phase. 
		We have to prove that this activation of $a$ links it into its correct place on the main branch. 

 The action of marking an address 
is part of the $\REMOVE(x)$ protocol (instruction at line 6).
After being marked, the marked address is still on the main branch, until an execution of instruction $\pred.\next:= d$ (line 2.7)
 physically removes it and thus ``cleans'' the main branch and makes room for the possibility of another $\ADD(x)$ execution. 
Marking an address is the only way by which the \ADD\ and \REMOVE\ protocols can find out that this address (after its removal) is not on the main branch 
(the Success Condition fails on a removed address but it may fail also on a marked address that is still on the main branch).
Without marking, an \ADD\ operation can activate an address on a secondary branch which is not what it is suppose to do, and a \REMOVE\
operation may fail to remove an address. 

We assume that, initially, for every address $a$
other than \head\ and \Tail\ $a.marked=\FALSE$ and
$a.\next =\bot$. An active address is an address that is either marked or on the main branch. Note that
\PActive\ is a predicate that will help in the correctness proof, but there is no address field that tells if an address is or is not active.    
An important property of active addresses is that  if $a$ is an active address, different from \Tail, and if $b=a.\next$, then $b$
is active and
$a.\kval < b.\kval$. This property holds even if $a$ is marked or removed.

Both the \ADD\ and \REMOVE\ protocols start by calling the $\locate(x)$
procedure which returns two active addresses \pred\ and \curr\ such that 
$\pred.\kval < x \leq\curr.\kval$ and (at the time when \curr\ is defined) $\curr= \pred.\next$. 
A possible problem however with these addresses is that, by the time $\locate(x)$ returns to the calling process,  relation $\curr =\pred.\next$ 
may fail.
For this reason, the calling protocol (\ADD\ or \REMOVE) locks the address \pred\ and then checks the success condition
that $\pred.\next= \curr$ and that $\pred$ is unmarked\footnote{It is conceivable that between the assignment
that determines \curr\ and the successful checking there was a moment when $\pred.\next\neq \curr$.}. We shall prove
that  the success condition implies that \curr\ is also unmarked.


A {\em global state} of the system is a description of the system at a certain moment. This description can be presented as a function
that assigns to every system variable a value in the domain of that variable.  We prefer to describe the state of the system
 at moment $t$ not as a list of values but rather as an abstract
structure (in the model theoretic sense).
The advantage of a structure over a plain list of values is in
that we have at our disposal a formalism for expressing properties
of structures in a precise and clear way. (Every structure
is related to some formal language that it interprets, and properties of that structure can be expressed in that language.)
The structures that replace here the notion of global state of the Lazy Set algorithm are called {\em Lazy Set state structures} and they
are defined in the following section.

\subsection{State structures and history sequences}
\label{Sec5}
In Definition \ref{Def4.1} we defined state structures for the Simpler Lazy Set algorithm (interpretations of the \SimplerLSstate) with which we represented
executions of the Lazy algorithm, and in this section we define a richer language ($\LSstate$) with structures that represent states of the full Lazy  Set
algorithm.

  As any logical language, 
$\LSstate$ is defined as a collection of symbols which are the names of sorts, predicates (names of relations), function symbols,
and names of members of the universe. 
\begin{enumerate}
\item The sorts are \Address, \Number, and $\cal P$ (for the set of processes), and the sort of instructions. Since as we can replace the
processes with id numbers and the instructions with line numbers,  sort $\Number$ can take care of processes and instructions.
The binary relation symbol $<$ is used as the ordering of the \Number\ sort.

\item  There are two predicates  defined over the address sort $\Address$: \PActive\ and  \PMarked.  

\item The functions are $\PNext:\Address \to \A$ , $\Val:\A\to \Number$, and $\PLockedto:\A\to {\cal P}\cup \{\bot\}$.

\item  A special constant $\bot$ is for the ``undefined''
value: for any function $f$, $f(x)=\bot$ means that $f$ is not defined at $x$.

\item 
The two program constants $\HH$ and $\T$ refer to  addresses.
For every process $P$, each of its local variables, $\pred_P$, $\curr_P$, and $\entry_P$  is also a {\em name} of an address in the address language
\footnote{In the terminology of model theory
$\pred_P$ is a constant of the language--in contrast to variables that can be quantified (as in $\forall x \varphi$).
But in our context, the term {\em program variable} would be more appropriate.}.
Other program variables in the address language are $x_P$ (of type $\Nat$) and the program counter $PC_P$ (of sort {\em instruction line} which is a subset of $\Nat$).

Note that the fields \next, \kval, and \marked\ of an address $a$ are not variables that the structure represents in a direct way.
Instead of interpreting $a.\marked$ we have the predicate $\PMarked(a)$; instead of interpreting $a.\kval$ we have $\Val(a)$ (a value in
$\Nat$) and instead of interpreting $a.\next$  we have $\PNext(a)\in \Address$.

\end{enumerate}

\begin{definition}
\label{DEF5.1}
A {\em state structure} $S$ is an interpretation of $\LSstate$; it consists of the following items. 
\end{definition}
 \begin{enumerate}
\item
A fixed set of addresses, $A$,  is the interpretation of the address sort in any
address structure: $A=\Address^S$  is a fixed, countably infinite set. The sort \Number\ has also
a fixed interpretation, that is $\Nat\cup \{-1,\infty\}$. The sort of processes is interpreted as a finite subset of $\Nat$. 

\item
The names,
 $\HH$ and $\T$ are fixed addresses (program constants) in $A$ (they have the same values in every address state).

\item
For every address predicate $R$, $R^{S}\subseteq A$ is the interpretation of $R$.
For every function symbol $F$, $F^{S}$ is a function defined over $A$ (or a subset of $A$). 
For example, $\PMarked^S$ is a subset of $A$, and $\PNext^S$ is a function from $A$ to $A$.

\item For every address $a\in A$, $\Val^S(a)\in \Nat \cup \{ -1, \infty\}$ (the key value of $a$),
and $\PNext^S(a)\in\Address$.  
 $\
\Val(\HH)=-1$, and $\Val(\T)=\infty$.

\item Finally,  every local program variable $v$ of a process has a value $v^{S}$ assigned to it
in  its type. Each local variable of process $P$ has $P$ as a subscript to differentiate it from the
variables with the same name that are local to the other processes. This being said, we may occasionally prefer to write \curr\ for example 
instead of $\curr_P$. 
(Unlike $\HH$ and $\T$, these names
 may designate different addresses in different structures.) 
\end{enumerate} 
 
 For every address $a\in A$, if $\PLockedto^S(a)=P$  then we say that address $a$ is 
locked by process $P$ in $S$.  $\PLockedto^S(a)=\bot$ means that $a$ is unlocked. We say that the key value of address 
$a$ is $v$ when $\Val^S(a)=v$. That is,
we use English instead of formulas when a more communicative tone is preferred.

\begin{definition}
\label{Def5.2}
{\bf The initial state structure} $\cal I$ is defined as follows.
\end{definition} For the predicates: \PActive\ applies only to
\HH\ and \T. Thus all other addresses are reserved for activation actions.
$\PMarked^{\cal I} =  \emptyset$. 
 For the functions: $\PNext(\HH)=\T$, $\PNext(\T)=\bot$, and $\PNext(a)=a$ for every other $a\in A$.
$\Val(\HH)=-1$, $\Val(\T)=\infty$, and $\Val(a)=\bot$ for every other address. $\PLockedto(a)=\bot$.
For the local variables, for every process $P$: $\pred_P^{\cal I} = \HH$, $\curr_P^{\cal I}=\T$, $\entry_P^{\cal I} =\bot$, and 
$PC_P^{\cal I}=0$. 

State structures are arbitrary interpretation of the $\LSstate$ language and may thus be quite far from what we expect them to be.
For example, in an arbitrary state structure there can be two addresses $a$ and $b$ such that $\PNext(a)=b\wedge \PNext(b)=a$ which
is never the case in executions of the algorithm.  We defined (in \ref{DefNS1}) normal states of the \SimplerLSstate\ language, and now we define normal \LSstate\ state structures.

\begin{definition}
\label{DPath}

A {\em path} in a state structure $S$ is a finite sequence of distinct addresses $a_1,\ldots,a_n$ such that $n>1$ and for every
$i<n$,
\[ a_i\neq \Tail\wedge a_{i+1}\neq a_i\wedge \PNext^S(a_i)=a_{i+1}.\] We say that this path leads from $a_1$ to $a_n$. Each of $a_1,\ldots,a_n$
is said to be on the path.
\end{definition}

\begin{definition} [Normal state structure.]
\label{DefAT}
A normal state structure is a state structure $S$ for \LSstate\ for which the following  conditions hold for every process $P$:
\end{definition}
 
\begin{enumerate}
 
\item[NS1] \begin{enumerate}    
\item \HH\ and \T\ are active addresses. $\Val(\HH)=-1$ and $\Val(\T)=\infty$.

 \item If $a$ is an active address that is different from $\HH$ and $\T$, then $\Val(a)\in \Nat$. If $a$ is active  and different from $\T$ then $\PNext(a)$ is active,
 and $\Val(a) < \Val (\PNext(a))$. 

\item
For every process $P$, $\pred_P$ and $\curr_P$ are active.

\item There is only a finite number of active addresses.

Also, the relation $a>b$ iff there is a path from $a$ to $b$ is a tree order on the set of active
addresses in which $\T$ is the root. I.e. there are no loops in this relation.)

\end{enumerate}
\item[NS2]
For every address $a$, $\PActive(a)$ iff $a$ is on the main branch or $\PMarked(a)$.

\item[NS3] Concerning the $\locate(x)$ procedure:
\begin{enumerate}

\item If $PC_P=2,3.1$ then $\Val(\curr_P)<x_P$ ($2$ and $3.1$ are line numbers in the \locate\ procedure. 

\item If $PC_P=3.2$ then $\pred_P=\curr_P$ and $\Val(\pred_P)<x_P$. 

\item If $PC_P = 4$ then $\Val(\pred_P)<\Val(\curr_P)$ and $\Val(\pred_P)<x_P$.
\item If $PC_P =5$, then $\Val(\pred_P)< x_P \leq \Val(\curr_P) $ 
\end{enumerate}

\item[NS4] Concerning the $\ADD(x)$ protocol:
\begin{enumerate}
\item If $PC_P= 1.2\ to\ 1.4$, then $\Val(\pred_P)< x_P \leq \Val(\curr_P)$.
If $PC_P=1.5$ then $\Val(\pred_P)< x_P < \Val(\curr_P) $.

\item If $PC_P=1.3$ to $1.6$, then $\PLockedto(\pred_P)=P$.

\item If $PC_P =1.4,1.5$ then $\pred_P$ is unmarked and $\PNext(\pred_P)=\curr_P$.

\end{enumerate}

\item[NS5] Concerning the $\REMOVE(x)$ protocol:
\begin{enumerate}
\item 
If $PC_P =2.2,2.3,2.4$, then $\Val(\pred_P)< x_P \leq \Val(\curr_P) $ 
\item
If $PC_P=2.5$ to $2.8$, then $\Val(\pred_P)< x_P = \Val(\curr_P) $.

\item If $PC_P= 2.3$ to $2.8$, then $\PLockedto(\pred_P)=P$. And if
$PC_P=2.6$ to $2.8$, then $\PLockedto(\curr_P)=P$. 

\item If $PC_P=2.4$ to $2.8$, then $\neg\PMarked(\pred_P)$.
If $PC_P=2.4$ to $2.7$ then $\PNext(\pred_P)=\curr$.
\item If $PC_P=2.7$ then $\PMarked(\curr_P)$.
\end{enumerate}

\end{enumerate}
It follows from these NS1 that if $a$ is active and different from \T, then
there is a path from $a$ to \T. In particular there is a path that leads from $\HH$ to $\T$. It is called the ``main branch'' of $S$.
Addresses \HH\ and \T\ are on the main branch,
they are active and unmarked. $\PNext(\T)=\bot$.
In Theorem \ref{ThmNormality} we shall prove that every state in an execution of the Lazy Set algorithm is normal.
We make some notes on this definition.
\begin{enumerate}
\item The initial state structure $\mathcal I$ is normal. Its only active addresses are $\HH$ and $\T$, and its main branch is
the path $(\HH,\T)$.
 $\pred_P$ and $\curr_P$
are active addresses in the initial state since they point to the \head\ and \Tail.
 
\item If there is a path from $a$ to $b$ then there is a unique such path (since \PNext\ is a function). Moreover, if \T\ is on that path then
$b=\T$ (as $\PNext(\T)=\bot$). 

\item If there is a path from $a$ to $b$ in a normal state then there is no path from $b$ to $a$. 
(Since in this case $\Val(a) < \Val(b)$.)
\end{enumerate}

We next define the notion   \PRemoved.
\begin{definition}
\label{DefForm}

For every address $a$ different from \HH, $\PRemoved(a)$ iff  $a$ is marked but there is no path from \HH\ to $a$. 
\end{definition}

\subsubsection{Execution triples and history sequences}
  \label{SETA}

An action by process $P$ is an atomic execution of an instruction of the protocol that $P$ is executing.
We are using state structures  (rather than tables of values), and we have to define how actions change these structures. We shall define the triples  
$(S,e,T)$ (also called steps) of state structures $S,T$ and action $e$ such that  $e$ is enabled at $S$ and transforms state $S$ into $T$. (We shall also
define predicates and functions 
on the actions  will which will help us later in section \ref{Sec6}.)
With this definition of steps as triples
we shall be able to define histories of the Lazy Set algorithm as sequences $S_0,e_0,S_1,\ldots,S_i,e_i,S_{i+1},\ldots$
such that $S_0$ is the initial address state, and for every $i$ $( S_i,e_i,S_{i+1})$ is an execution triple by one of the processes. 

\begin{definition}
\label{D5.5}
We say that a pair of states $(S_1,S_2)$ is an $(i_1,i_2)$ step by process $p$
 if for some event $e$ $(S_1,e,S_2)$ is a triple as defined below for process $p$ such that 
$PC_P^{S_1} = i_1$ and $PC_P^{S_2}=i_2$.
\end{definition}
Some examples. A $(0,3.1)$ step is a \CONTAINS\ invocation, and a $(3.5, 0)$ is its returning response; a $(1.1,1)$ step is a $Locate$ invocation, and a $(5,1.2)$ is its response (and assignment to $(\pred,\curr)$); a $(1.5,1.6)$ step is an activation, and a $(2.7,2.8)$ step
is a physical removal.

\noindent
{\em Definition of the triples (also called steps)} 
\begin{enumerate}
\item Invocation and response triples  are defined with reference to the generic algorithm of Figure \ref{F6}.
The triple $(S,e,T)$ is an invocation of $\ADD(x)$ by process $P$, for example, when the program counter $PC_P$ of $P$ is on
line $0$ of the generic algorithm 
and is on the first line of the $\ADD(x)$ protocol in $T$ with $x_P$ an arbitrary value in $\Nat$. The triple $(S,e,T)$ is a corresponding 
response when $PC_P$ is a
return instruction in $S$ and is back to line $0$ in $T$.  (There are three possibilities for an \ADD\ execution to return:
at line 1.3 (when failed), at line 1.4 (returning $1$), and at 1.7 (returning $0$).

\item
$(S,e,T)$ is a {\em locking} triple by process $P$ when $e$ is an execution of an instruction of the form $v.lock()$ where $v$ is an address variable
of the process $P$ that executes this action. Suppose that $a=v^S$ is the address that $v$ refers to at address structure $S$.
 Action $e$ is enabled at $S$ when $\PLockedto^S(a)=\bot$ (meaning that $a$ is unlocked in $S$) and $PC_P^S$ is a line with that locking instruction. In this case
we have in $T$ that $\PLockedto^T(a)=P$, and $PC_P$ refers to the next instruction of the protocol. In all other features, 
$T$ is equal to $S$. While  $\PLockedto(a)=P$, no process can execute a locking step and thus the intervals formed by locking and unlocking
of address $a$ are mutually exclusive. (The predicate $P$ applies to $e$ to say that $e$ is an action by $P$, the predicate {\em locking}
applies to $e$, and the function $\PLockedto$ is defined at $e$ with value $P$.)

An {\em unlocking} action $e$ by process $P$ is enabled at $S$ when $\PLockedto^S(a)=P$ and the program counter of $P$ refers to some instruction
of the form $v.\unlock()$ where $a=v^S$. Unlocking triples are naturally defined. (The predicates $P$ and {\em unlocking} are defined
on $e$, and the value $\PLockedto(e)=\bot$ is defined.)

\ignore{
\item An {\em instantiation} triple $(S,e,T)$ is an execution of  $\entry := \New(\kval:=x,\next:=\curr)$. 
The following conditions hold. 

\begin{enumerate}

\item
 $PC_P$ refers to line 1.5
of the \Add\ protocol in $S$ and in $T$ it refers to  line 1.6.
\item For some address $a$ such that $\PReserved^S(a)$, $\neg \PReserved^T(a)$, 
  and $\entry_P^T= a$. We say that action $e$ is an instantiation of address $a$.
	\end{enumerate}
	
	In all other
features $T$ equal $S$.  (The predicates $P$ and {\em instantiation} apply to $e$.)
}

\item      An {\em activation} triple is an execution of instruction $\pred.\next:=  \New(\kval:=x,\next:=\curr) $ at line 1.5 of the \ADD\ protocol.
The triple $(S,e,T)$ is an activation  when the following hold for some address $a$.
\begin{enumerate}
\item $PC^S_P= 1.5$. $\neg \PActive^S(a)$.

\item $PC_P^T= 1.6$. $\Val^T(a)=x$, $\PActive^T(a)$ and \[ \PNext^T(\pred^S)=a \wedge \PNext^T(a)=\curr^S.\] 
 There are no other changes and in particular, for all addresses $d\neq \pred^S$,
 $\PNext^T(d)=\PNext^S(d)$.

Note that once an address is activated it stays active forever because there is no action that returns it to be non-active. 
  This ensures that no address $a$ is activated  twice.)
 \end{enumerate}

\item A {\em marking} triple $(S,e,T)$ is an execution of instruction $\curr.\marked := \TRUE$ (line 2.6) by
 process $P$. A marking action is enabled at $S$ when $\PLockedto(\curr^S )=P$ and the program counter of $P$ refers
to that marking instruction. As a result of the marking action we have that $\PMarked^T(\curr^S)$ holds.
We say that $e$ is a {\em marking of address} $a=\curr^S$ (and the predicate {\em marking} is applied to action $e$).

\item A {\em physical removal} triple $(S,e,T)$ is an execution  of the  $\pred.\next:= \curr.\next$ instruction in the 
$\REMOVE(x)$ operation. We have $PC_P^S= 2.7$, $PC_P^T= 2.8$, and
\[ \PNext^T(\pred^S)=\PNext^S(\curr^S).\] We say that $e$ ``physically removes address  $a= \curr^S$.

\item Reading actions  change the values of local program variables. A reading action does not require any locking rights.
 When $(S,e,T)$ is a reading action we say that ``$e$ reads from $S$''. For example, if
$e$ is an execution of $v:=\curr.\kval$ we define $\kval(e)=\Val^S(\curr^S)$. 

\item Although it occupies just one line, an execution of $(\pred,\curr):=\locate(x)$ is clearly not an atomic action. It consists
of an atomic call to the \locate\ procedure, a series of read actions, and a return action. 

\item $(S,e,T)$ is a $(1.4,0)$ triple by process $p$ 
 when $PC_p^S=1.4$, $\PLockedto^S(\pred_p)=p$, and $\Val(\curr_P)=x_p$ holds in $S$. 
Then $PC_p^T=0$, $\PLockedto^T(\pred_p^S)=\bot$, and $e$ is a return event such that $\chi(e)=1$. Now, 
$(S,e,T)$ is a $(1.4,1.5)$
triple when  $PC_p^S=1.4$ and $\Val(\curr_P)\neq x_p$. In this case $PC_p^T=1.5$ is the only change. 
\item $(S,e,T)$ is a $(2.4,0)$ triple by process $p$ when $PC_p^S=2.4$, $\PLockedto^S(\pred_p)=p$,
and $\Val(\curr_p)\not> x_p$. In this case $\PLockedto^T(\pred_p^S)=\bot$ and $PC_p^T=0$.
$e$ is a return action with $\chi(e)=0$.
\end{enumerate}

\begin{lemma}
\label{L3.3}
The following holds for every execution triple $(S_1,e,S_2)$.\\  (1) $\HH^{S_2} = \HH^{S_1}$ and $\T^{S_2}=\T^{S_1}$. (2)
 For every address $a$, if $a$ is \PMarked\ at $S_1$ then it remains marked at $S_2$, and (3)  If $a$ is active at $S_1$ then it remains active at $S_2$.
\end{lemma}
Proof. Going over the actions it is easy to check that no action changes the interpretation of $\HH$ or $\T$, and no action
can unmark an address. Likewise no action changes an address from being marked to unmarked. \qed

Executions of the Lazy Set algorithm can be presented by means of history sequences in which actions by the different processes are
interleaved.
\begin{definition}[History sequence of state structures]
\label{DHS} 
A history sequence is a sequence $H= (S_0,e_0,S_1,\ldots,\ldots,S_i,e_i,S_{i+1},\ldots)$ (finite or infinite) such 
that $S_0$ is the initial state structure {$\mathcal I$}, and for every $i$ $(S_i,e_i,S_{i+1})$ is an execution triple as defined above by one of the processes. As usual, the fairness condition is that in any infinite history sequence each process should have
an infinite opportunity for actions. 
\end{definition}

\begin{theorem}
\label{ThmNormality}
In any history sequence of the Lazy Set algorithm every state is normal.
\end{theorem}
For the proof we let ``normality'' be the conjunction $\varphi$ of the statement NS1 to NS5
and prove that $\varphi$ is an invariant. It is simple to check that the initial state is normal, and
the main part is the proof that for every triple $(S,e,T)$, if $S$ 
is normal then $T$ is normal. This is rather standard and long, and hence instead of a detailed proof we outline an informal argument as an evidence that every state in a history is normal. 
\begin{enumerate}
\item For the proof of NS1, we note that \HH\ and \T\ are active in the initial state (Definition
\ref{Def5.2}) and their values is $-1$ and $\infty$. No action can deactivate an address or changes
its value. This establishes NS1(a).

Only activation action (i.e. $(1.5,1.6)$ actions $(S,e,T)$) can change the \PActive\ predicate
on an address from inactive to active. The new address $\adr(e)=a$ that is activated has a value
in $\Nat$. Since $S$ is normal and $PC_P^S=1.5$, $NS4(a),(c)$ hold in $S$, and hence
$\Val(\pred_P)<x_P<\Val(\curr_P)$ and $\PNext(\pred_P)=\curr_P$ hold in $S$. 
By definition of activation triples, $\PNext(\pred_P^S)=a$ and $\PNext(a)=\curr_P^S$ hold in $T$.
This implies that $a$ is introduced in its place in the main branch of $T$. So NS1 holds in $T$,
and more generally $T$ is normal. (The number of active addresses is increased by one, and remains
finite of course.) 

\item The proof of NS2 depends on the observation that any physical removal triple of address $a$
is preceded by a marking step of $a$ (formally statement NS5(e) says that if $PC_P=2.7$ then
$\curr_P$ is marked).

\item For the proof of NS3, note first that $\pred_P$ and $\curr_P$ are initially \HH\ and \T, so 
that they are initially active. The statements (b) to (e) of NS3 refer only to variables that are
local to process $P$, and no other process can change them. Therefore it suffices to prove
these statement under the assumption that process $P$ executes \locate\ with no interleaving
with actions of other processes. 

\item Item NS4(a) is basically a consequence of NS3(e). For NS4(c) we note that steps
$(1.3,1.4)$, $(1.4,1.5)$, and $(1.5,1.6)$ are executed while $\pred_P$ is locked to $P$.
This implies that only process $P$ can change any field of $\pred_P$. Since process $P$
reaches $PC_P=1.4$ after the success condition is established, $\neg\PMarked(\pred_P)\wedge 
\PNext(\pred_P)=\curr_P$ (that is NS4(c)) holds at any state between the locking and unlocking 
of $\pred-p$. 
\item NS5 is argued in a similar fashion. 
\end{enumerate}


\subsection{Linearizability of the Lazy Set algorithm}
\label{Sec6}

In this section we define the reducts of \LSstate\ structures to \SimplerLSstate\ structures and explain how this can lead to a proof of linearizability of the Lazy Set algorithm. First we define reducts in general, then we fine-tune this definition to the case that interest us 
(\LSstate\ to \SimplerLSstate) and state the properties of this reduct which directly yield the desired linearizability.

Let $L_1$ and $L_2$ be two languages with $L_2$ being richer than $L_1$. If $M$ is an $L_2$ structure, 
then $\pi M$, the reduct of $M$ to $L_1$, is
defined as the structure that interprets $L_1$ and is obtained by taking from $M$ only the interpretations of the symbols
of $L_1$. So the universe of $\pi M$ is obtained by taking from $M$ the interpretations of the sorts that are in $L_1$,
and if $R$ for example is a predicate in $L_1$ (and hence in $L_2$) over some sort $X$, then $R^{\pi M}=R^M$. Likewise, $\pi M$ interprets
the function symbols of $L_1$ the way $M$ does.

Fix in this section an infinite history sequence 
\[H= (S_0,e_0,S_1,\ldots,\ldots,S_i,e_i,S_{i+1},\ldots)\] as in Definition \ref{DHS}.
 We define the reduct (or projection) $\pi H$
of history $H$ by (essentially) looking at the sequence $\pi H = (\pi S_i\mid i=0,\ldots)$ of reducts to the \SimplerLSstate\ language. Below we shall give a detailed definition of $\pi H$, but first we want to describe the general idea
and how we are going to use $\pi H$ to get a proof of linearizability of the Lazy Set algorithm. For most indexes $i$, we shall have
$\pi S_i = \pi S_{i+1}$. That is, many of the steps (or triples) $(S_i,e_i,S_{i+1})$ of $H$ are reduced to stuttering
triples which do not change the reduced structures. Steps of $H$ whose reducts are meaningful non-stuttering
steps are for example the actions that change the graph structure of the states, that is the events that change the \PNext\ function
(namely the activate and physical removal steps).
A marking step for example induces a stuttering step because the reduct of a structure forgets the \PMarked\ predicate.
Likewise any step that is an execution of an instruction of the  $Locate$ procedure has to induce a stuttering step because
the Simpler algorithm has no such procedure. After defining the reduct of history $H$, we shall prove that it is a history sequence
(with stuttering steps) of the Simpler algorithm. Since we have already proved the linearizability of that algorithm, an immediate
proof of the full Lazy Set algorithm will be obtained.

If $S$ is a normal state of the \LSstate\ language, then we want to define its reduct, $\pi S$, as a is a state of the \SimplerLSstate\ language.
The universe of $\pi S$ is $A\cup {\Nat}\cup \{-1,\infty\}$ where $A$ is the fixed set of addresses (this is the universe of
any \SimplerLSstate\ structure). Then $\pi S$ has the same interpretation of the \PActive\ predicate, and the same \PNext\ and \Val\ functions. But $\pi S$ ``forgets'' about the \PMarked\  predicate and about the \PLockedto\ function. 
We have however a certain problem with the program counters $PC_p$, because the program counter of processes that executes the 
Lazy Set algorithm and those that execute the Simpler
algorithm have different types: the first gets values that are lines of the former algorithm and the second of the latter.
Hence the definition of the reduct function $\pi$ that reduces $\LSstate$  to $\SimplerLSstate$ structures has to take this difference
into acount (for we want that $\pi S_i$ be a state of the Simpler algorithm). Let $\SimplerLSstate^-$ be
the language obtained from $\SimplerLSstate$ by removing all the $PC_p$ constants. Given a state $S$ for
the $\LSstate$ language, we say that the reduct of $S$ to $\SimplerLSstate^-$ is a {\em partial reduct}. The partial reduct of a state $S$ is denoted $S^-$.
 The full reduct of $S$, denoted $S'$,  is obtained from the partial reduct $S^-$ by specifying the value of the program counters.
The initial idea for this  is to define a function, still called $\pi$, from the set of lines  of the Lazy Set algorithm to the set of lines
of the Simpler algorithms, and then to define $S'$ by specifying that $PC_p^{S'} =\pi(PC_P^S)$. There is a wrinkle that has to be ironed
out before we can proceed with this idea.

Take for example a $(1.1,1)$ triple $(S_i,e_i,S_{i+1})$ in history $H$,
that is a call to the $\locate(x)$ procedure by process $p$ that executes line 1.1 of the \ADD\ protocol. 
So $PC_p^{S_i}=1.1$ and $PC_p^{S_{i+1}}=1$. We assume that $S_i'$, the
reduct of $S_i$, is defined and is such that $PC_p^{S_i'}=1$. Our aim is to define $S_{i+1}'$ in such a way that $(S_i',S_{i+1}')$
is a stuttering step (knowing that the execution of the Simpler algorithm is not aware of the $\locate$ procedure). For this, we define
$PC_p^{S_{i+1}'}=1$ and observe that $S_i'=S_{i+1}'$. If, however, $(S_i,e_i,S_{i+1})$ is a $(2.1,1)$ step, namely a call to the 
same \locate\ procedure but now from the \REMOVE\ protocol, then we have to define $PC_p^{S_{i+1}'}=2$ (rather than $1$). 
The conclusion is the for lines $\ell$ of the \locate\ procedure 
the value $\pi(\ell)$ has to be $1$ or $2$ depending on whether the call of this procedure is from line $1.1$ or $2.1$.	,
The inner actions of the \locate\ procedure are reduced to stuttering steps (except for the response actions which are reduced to
responses that return to the caller).
	Granting this caveat, the definition of $\pi$ on the lines is in the following tables.

\begin{center}
\begin{tabular}{|c|c|c|c|c|c|c|c|c|c|c|c|c|c|c| } 
 \hline
 $\ell$  \      & 0 & 1.1& 1.2&1.3  &  1.4  & 1.5 & 1.6  & Any $Locate$ lines   \hspace{3mm}           \\
\hline 
 $\pi(\ell)$ \  & 0 & $1$& $1$& $1$ & 1     & 1   &  $0$ &  $1/2$ depending on the caller \\ 
\hline 
 \end{tabular}
\end{center}

\begin{center}
\begin{tabular}{|c|c|c|c|c|c|c|c|c|c|c|c|c|c|c|} 
 \hline
 $\ell$  \      &2.1& 2.2& 2.3  & 2.4 & 2.5 & 2.6 & 2.7 & 2.8 & Any line $3.\ell$\\
\hline 
 $\pi(\ell)$ \  & $2$& $2$& $2$ & 2   & 2   & 2   & $2$ & 0 & $3.\ell$ \\ 
 \hline
\end{tabular}
\end{center}

We shall illustrate now this choice of $\pi$ by describing a few examples of reducts of triples of the Lazy Set algorithm. 
 
\begin{enumerate}

\item  
A $(1.4,0)$ step $(S,e,T)$ by process $p$ that executes its \ADD\ protocol is a response step that returns the status value $1$.  Its reduct is  a $(\pi(1.4),\pi(0))=(1,0)$ step, namely a response triple of the Simpler algorithm. (It is not difficult to check that indeed  this reduct is a response triple.)
Any  $(1.4,1.5)$ step $(S,e,T)$ however is reduced to a $(\pi(1.4, \pi(1.5))=(1,1)$ stuttering
step. We have to check that indeed $S^-=T^-$ in order to conclude that the reducts form a stuttering step of the Simpler algorithm.

\item In a similar way, reducts of the \REMOVE\ steps are obtained. Note that a $(2.3,0)$ step $(S,e,T)$ (in which status $f$ is returned)
is reduced to a response step with status $0$ (and is such that $S^-=T^-$, i.e. only $PC_p$ changes, whereas a $(2.3,2.4)$ step is reduced to a stuttering step (as $S'=T')$. 
A $(2.4,0)$ step is reduced to a $(2,0)$ response step (with status $0$). The $(2.4,2.5), (2.5,2.6)$,
and $(2.6,2.7)$ steps are reduced to $(2,2)$ stuttering steps. For example, if $(S,e,T)$ is a $(2.5,2.6)$ step (namely a locking step)
then $S'=T')$ because the reducts of $S$ and $T$ do not take notice of the change in the \PLockedto\ function.   But if $t =(S,e,T)$
is a $(2.7,2.8)$ step in which process $p$ removes an address $a$ of value $x$ from the main branch of $S$ then $t$ is reduced to the physical removal $(2,0)$ step
of the Simpler algorithm. To prove that this is indeed the case, Lemma \ref{LEM5.15} establishes that all the requirements for 
physical removal triples hold. That is, that there is indeed an address of value $x$ on the main branch of $S$ and that it is no longer
on the main branch of $T$.
\item The \CONTAINS\ protocol is the same for the Lazy Set and the Simpler protocols. Since $\pi$ is the identity on the line numbers
of that protocol, if $(S,e,T)$ is an execution of instruction $\curr:=\curr.\next$ then the reduct $(S',e',T')$ is also an execution
of that instruction.
\end{enumerate}

Linearizability of the Lazy Set algorithm is a direct consequence of the following.
\begin{theorem} 
\label{TM5.10}
If $H  = (S_0,e_0,S_1,\ldots,\ldots,S_i,e_i,S_{i+1},\ldots)$ is a history sequence of the Lazy Set algorithm, then the sequence of reducts of the states of $H$, $H'=(S_0',e_0',\ldots,S'_i,e'_i,S'_{i+1},\ldots)$ is a history sequence (with
stuttering steps) of the Simple Lazy Set algorithm. 
\end{theorem}
The proof of the theorem is by means of the following five lemmas. Then linearizability of $H'$ (a consequence of Theorem
\ref{MSimpler}) directly implies that $H$ is linearizable.

 \subsection{Five important lemmas}
Fix a history sequence $H  = (S_0,e_0,S_1,\ldots,\ldots,S_i,e_i,S_{i+1},\ldots)$ of the Lazy Set algorithm.
Our aim is to prove that the sequence of reducts of the states of $H$ generate a history sequence $H'$ of the Simpler algorithm. 
We start with the following easy lemma.
\begin{lemma}
Suppose that $S_i$ and $S_{i+1}$ are two successive states in the history sequence $H$ that describe a step by process $p$, and 
let $S_i'$ and $S_{i+1}'$ be their reducts.
If the program counter $PC_p$ has the same value in $S_i'$ as in $S_{i+1}'$ then $S_i=S_{i+1}$ (i.e. this is a stuttering step). 
\end{lemma}
Proof. We have to check all the possible triples $(S,e,T)$ of the Lazy Set algorithm, and
for each such triple for which $\pi(PC_P^S)=\pi(PC_P^T)$ we have to check that
$S'=T'$. Take for example a $(1.2,1.3)$ step $(S,e,T)$; since $\pi(1.2)=\pi(1.3)=1$, we have
to check that $S'=T'$. This is indeed so because $(S,e,T)$ is in this case a locking step and
the reducts of $S$ and $T$ forget about the \PLockedto\ function, so that $S'=T'$.
To give another example, consider a $(2.1,1)$ step $(S,e,T)$. This step is a call to the \locate\
procedure from the \REMOVE\ protocol. So $\pi(1)=2$ because the call to line $1$ of \locate\
is from line $2.1$ and $\pi(2.1)=2$. And indeed $S'=T'$ so that $(S',e',T')$ is a stuttering step.
We leave the completion of the proof of this lemma to the reader. \qed

Steps in $H$ that are executions of some $\CONTAINS(x)$ instruction are reduced to the same steps of the corresponding protocol
of the simpler algorithm. The invocation steps in $H$ are reduced to invocation steps in $H'$. 

An activation action of the \ADD\ protocol is a $(1.5,1.6)$ step. The following lemma establishes that the reduct of this action
is a $(1,0)$ step of the Simpler algorithm, i.e. an $\AD(x)$ execution.

\begin{lemma}
\label{L5.14}
Suppose that $S$ is a normal state and $(S,e,T)$ is an activation action in some $\ADD(x)$ execution $E$ by process $P$, i.e. an execution of a $(1.5,1.6)$ step in $E$. Let $a$ be the address that is activated by this step. 
Then the reduct triple $(S',e',T')$ is a $(1,0)$ $\AD(x)$ action that adds $a$ to the main branch of $S$.
\end{lemma}
Proof. 
Since $PC_P^S=1.5$, $S$ satisfies the following properties.
$\pred_P$ and $\curr_P$ are active addresses (by NS1(d)), and
\begin{equation}
\label{EM} \Val(\pred_P)<x_P<\Val(\curr_P)\end{equation}
(by NS4(a)). Also, by NS4(c), $\neg \PMarked(\pred_P)\wedge \PNext(\pred_P)=\curr_P$. So $\pred_P$, 
being active and unmarked,
is on the main branch of $S_i$ (by NS2), and hence $\curr_P$ is also on that main branch (by the
 definition of the main branch and as $\PNext(\pred_P)=\curr_P$). Since $S$ is a normal state,
 Equation (\ref{EM}) implies that there is
no address of value $x$ on its main branch. In $T$, $a$ is an address of value $x$ inserted at the right place (by the definition of this step). Hence $(S_i',e_i',S_{i+1}')$ is a $(1,0)$ $\AD(x)$ action. \qed

The following lemma considers a $(2.7,2.8)$ physical removal triple in $H$ and proves that its 
reduction is a $(2,0)$  \RM\ action.

\begin{lemma}
\label{LEM5.15}
Suppose that $S$ is a normal state and  $(S,e,T)$ is a $(2.7,2.8)$ physical removal triple in some $\REMOVE(x)$ operation execution $R$ by process $P$ that
removes address $a=\curr^{S}$. Then the following hold.
\begin{enumerate}
\item[In $S$:] $a=\curr^{S}$ is an address of value $x_P$ on the main branch, and $\PNext(\pred)=\curr$ holds in $S$.
 Say $p=\pred^{S}$
and $d=\PNext^{S_i}(a)$.

\item[In $T$:] $\PNext(p)=d$. So address $a$ is no longer on the main branch of $T$.
\end{enumerate}
\end{lemma}
Proof.
In $S$, $PC_P=2.7$ and so we have the following in $S$.
\begin{enumerate}
\item $\PLockedto(\pred_P)=P$ and $\PLockedto(\curr_P)=P$ (by NS5(c)).
\item $\PMarked(\curr_P)$ (by NS5(e)).
\item $\Val(\curr_P)=x_P$ (by NS5(b)).
\item $\neg\PMarked(\pred_P)$ (by NS5(d). So, since $\pred_P$ is active, $\pred_P$ is on the main branch.
\item $\PNext(\pred_P)=\curr_P$ (by NS5(d)). So $\curr_P$ is also on the main branch.
\end{enumerate}
Say $p=\pred^{S}$ and $d=\PNext^{S}(a)$.
since $(S,e,T)$ is a removal step, $\PNext^{T}(p)=d$. Thus $S',T')$ is indeed a
 $(2,0)$ removal action of the Simpler algorithm. 
 \qed

The following lemma deals with \ADD\ operation of status $1$. We have to show that the reduct of any 
$(1.4,0)$  action
of the Lazy Set algorithm is a $(1,0)$ \AD\ action of status $1$.

\begin{lemma}
\label{LEM5.13}     Suppose that $S$ is a normal state and
 $(S,e,T)$ is a $(1.4,0)$ action by process $P$ (which returns status $1$).
 Say $a=\curr^{S}$.
 Then in states $S$  address $a$ is on the main  branch and
with value $x$. Thus the reduct $(s',e',T')$ is a $(1,0)$ step
\end{lemma}
Proof. 
In $S$, $PC_P=1.4$, and in $T$, $PC_P=0$. As specified in the lemma, $a=\curr_P^{S}$.
 Since $S,e,T)$ is a $(1.4,0)$ step, we have by definition of that
step that $\Val(a)=x_P$ holds in $S$. We have to prove that $a$ is on the main branch of $S$ in
order to conclude that $(S',T')$ is a $(1,0)$ step of the Simpler algorithm. 
Now, $\pred_P$ is unmarked in $S$ and $\PNext(\pred_P)=\curr_P$ holds in $S$ (by NS4(c)).
Hence (being unmarked and active) $\pred_P^{S}$ is on the main branch of $S$, and so
 $\PNext^{S}(\pred_P^{S})=\curr_P^{S}$
is also on the main branch of $S$. That is, $a$, of value $x_P$, is on that main branch. \qed

The fifth lemma deals with a $(2.4,0)$ action and proves that its reduct is a $(2,0)$ action
of the Simpler algorithm.

\begin{lemma}
Suppose that $S$ is a normal state and $(S,e,T)$ is a $(2.4,0)$ triple (so $\chi(e)=0$). Then $(S',e',T')$ is a $(2,0)$ action
of the Simpler algorithm.
\end{lemma}
Proof. Since $PC_P^S=2.4$ and $PC_P^T=0$, the definition of the \IF\ statement at line 1.4
implies that $\Val^S(\curr_P)=x_P$. Condition NS5 implies that the following hold in state $S$.
\[ \neg \PMarked(\pred_P)\wedge \PNext(\pred_P)=\curr_P.\]
Hence $\pred_P^S$ (being unmarked and active) is an address on the main branch of $S$. Hence
$\curr_P^S$ is also an address on the main branch of $S$, and its value is $x_P^S$.
This is the condition required for $(S',e',T')$ in order that it be a $(2,0)$ step of the Simpler
algorithm. \qed

\section{Conclusion}

We proved here the linearizability of a version of the Lazy Set algorithm of \cite{Hell1} with the aim for promoting
an approach which is not based on searching for linearization points, but is rather concerned, in a large part,
with abstract high level properties that imply linearizability. These properties (the Lazy Set axioms of Section \ref{Sec2}) are expressed
in some mainly first-order language, and a theorem is proved in that section 
(that is Theorem \ref{MT}) that any structure that satisfies
these axioms is linearizable.  In order to apply this approach we must be able to transform executions of the protocol
into Tarskian structures, namely interpretations in the model theory sense, for which it is meaningful to say that
they satisfy (or not) the sentences that form the Lazy Set axioms. For this aim, we have to start with the usual definition
of executions as history sequences and to define for each history $H$ that describes a run of the Lazy Set algorithm a structure
$M_H$ that corresponds to $H$. In fact, it is not difficult to see that it is possible to retrieve $H$ from $M_H$, so that
these two mathematical objects are closely related. For this reason it should be possible in less than formal 
proofs to argue directly about histories and to prove, at least intuitively, that any history that is a run of the algorithm
satisfies the axioms. (I recommend however that the formal language in which the axioms are expressed is carefully defined; here an informal language may lead to unreliable proofs.) In fact, our development of the linearization proof of the Lazy Set algorithm was not so direct: for we presented a shorter and simpler algorithm, the Simpler Lazy Set algorithm, for which
we thought that it would be easier to define the Tarskian structures $M_H$ that correspond to history sequences of that
Simpler algorithm. Then we proved that every such structure $M_H$ that describes an execution of the Simpler algorithm satisfies the Lazy Set axioms and is thence linearizable. In this fashion the proof of linearization for the Simpler
algorithm is obtained (Section \ref{MSimpler}). Luckily, for the proof of the fuller Lazy Set algorithm, there is no need
to repeat this work again. It suffices to observe that any execution of the fuller algorithm has a projection, a reduct,
that is an execution of the Simpler algorithm. This observation necessitates a proof of course, which is quite
straightforward    (see Theorem \ref{TM5.10}).

\subsubsection{A note on this version of the Lazy Set Algorithm}
\label{Variants}

There are several published versions of the Lazy Set algorithm, and in fact even the inventors of the algorithm present two
versions that are slightly different: whereas the version presented in \cite{Hell1}
needs two locking actions in each Add operation, the one in \cite{Helleretal}  needs just one (two locking actions are still
needed for the Remove operations and none for the Contains operation). Comparing the algorithm presented here and the different
published versions (the original ones, papers mentioned in  \cite{DD}  and the presentation in \cite{HS}) we only note those which seem to me more important.

 A difference of lesser importance is that our \ADD\ and
\Remove\ operations may return the \Failed\ token which the corresponding operations in the original and all the other
versions never do. The point is that in these other versions the \ADD\ and \Remove\ operations begin with a  \WHILE(\TRUE)\ loop which
is possibly non-terminating, whereas we prefer to have only terminating procedures. In case our \ADD\ or \Remove\ operation execution
return the \Failed\ value (which as was explained in Section \ref{STS} is a normal condition), the calling procedure may decide to recall that operation as many times as needed (thus imitating
the \WHILE\ loop) or else may decide to invoke another operation. 
A related stylistic difference is that our operations return the values $1$ and $0$
which do not indicate success or failure as in the other versions, but rather whether an address with value
$x$  is or is not on the set.

More interesting differences show the concern for leaner operations. For example we try to minimize the usage of locking instructions.
Our \Remove\ operation needs only one locking, the locking of address \pred, in case that the value $0$ is returned (locking two addresses
is needed only for successful \Remove\ operations--those that return $1$). In all other versions of the
algorithm two lockings are needed in every \Remove\ operation both when returning \FALSE\ and returning \TRUE. 
The \ADD\ protocol needs only one locking in \cite{Helleretal} and in our paper, but it needs two locking (of \pred\ and \curr) 
in the earlier version \cite{Hell1}, in \cite{FV}, \cite{DD}, \cite{CA} (these papers use two locking for the \Remove\ operations
as well).

It is in the \Contain\ procedure and the fact that it is wait-free (as it does not use any locking) that lies the most important
contribution of the Lazy Set algorithm. It is also that part of the algorithm that creates the greater difficulties in the
correctness proofs.  It is here also that lies also a somewhat surprising difference between the version presented here and the
all the other published versions. In all of these versions a process executing $\Contain(x)$ traverses the list (disregarding locks) from Head towards the Tail, stopping as soon as a node with value greater or equal to $x$ is found. Value true is returned
if the node is unmarked and its value is equal to x, otherwise false is returned. The code for deciding the value to
return is the following.
 \begin{equation}
\label{DIS} \text{return (!curr.mark) and (curr.val = x)}\end{equation}
(here ``!'' is the negation symbol). 
In our version of the $\Contain(x)$ procedure the code is shorter; the list is traversed in increasing order from the Head, until a node
with value greater or equal to $x$ is found, as in all the other versions, but the decision of the value to return is simpler:
\[ \text{return (curr.val = x)}.\]
That is, the $\Contain(x)$ code looks only at the value and next fields, and disregards both the locks and the marking fields.
I confess that even to me it looked strange to realize that the $\Contain(x)$ execution returns the value \TRUE\ (saying
that $x$ is in the set) even in the case that the node with key $x$ that it finds is a {\em marked} node, i.e. a node that is
marked for removal. 

The importance of this tighter version of the \Contain\ code is not in that it is more efficient (surely testing the
{\em marked} field takes practically no time), but rather in that the tighter version yields a deeper understanding of the
algorithm. Namely, it is evident now that the marking mechanism of the Lazy Set algorithm is not needed for the
\Contain\ procedure. (Although one may {\em prefer} to use the original \Contain\ version, this preference is not related to
the issue of correctness.)  
An interesting analysis of the code displayed in (\ref{DIS}) is in \cite{DD} (Section 5.1 there).  Dongol and Derrick
discuss the ordering of the two actions,  reading curr.mark and reading curr.val. They claim that both ordering is acceptable
but each of the proofs of correctness that they have found in the literature only addresses one of the alternatives, reading
first the mark field and reading first the value field. This is not an issue in our version.

\end{document}